\documentclass[
 amsmath,amssymb,
 aps,twocolumn
]{revtex4-2}
\usepackage{graphicx}
\usepackage{dcolumn}
\usepackage{bm}
\usepackage{color}
\usepackage{cancel}
\usepackage{booktabs}
\usepackage{comment}
\usepackage{braket}
\usepackage[colorlinks=true,
            linkcolor=blue,
            citecolor=blue,
            urlcolor=blue]{hyperref}

\begin{document}

\preprint{APS/123-QED}
\title{Thermoelectric information engine driven by an autonomous Maxwell demon across quantum-to-classical transitions}
\author{M. Bernal-Santibañez}
 \altaffiliation[]{maximiliano.bernal@ug.uchile.cl}

\author{José Mondaca }%
 \email{jose.mondaca.r@ug.uchile.cl}
\affiliation{%
 Departamento de Física, Facultad de Ciencias Físicas y Matemáticas, Universidad de Chile, 837.0415 Santiago, Chile
}%

\author{Felipe Barra}%
 \email{fbarra@dfi.uchile.cl}
\affiliation{%
 Departamento de Física, Facultad de Ciencias Físicas y Matemáticas, Universidad de Chile, 837.0415 Santiago, Chile
}%

\date{\today}

\begin{abstract}

We study a three-terminal thermoelectric engine, focusing on the role of quantum coherence and information flow. A double-dot connects two reservoirs at different chemical potentials, while a third dot monitors their occupation via Coulomb interaction and can be interpreted as an autonomous Maxwell demon. Within the parameter range where the device operates as an engine, we identify conditions under which this interpretation holds. Namely, when a finite information flow toward the monitoring dot is accompanied by negligible energy flow.
The system dynamics is described within a Redfield master equation that allows us to identify two distinct dynamical regimes with steady states well captured by suitable Lindblad approximations.
For large interdot tunneling, energy-basis coherences vanish, and the steady state reduces to that of a classical stochastic master equation associated with a fully secular approximation.
In contrast, for small tunneling, coherence persists in both energy and local eigenbases, and a Lindblad description derived from a partial secular approximation to the Redfield equation must be employed. 
These two regimes define a first quantum-to-classical transition controlled by the interdot tunneling strength. 
We further consider the effect of a phonon bath coupled to the double-dot, which induces a second quantum-to-classical transition by generating incoherent transport and decoherence in the small interdot tunneling regime. We find a competition between these effects: incoherent transport enhances both particle and information flows, whereas decoherence suppresses them, and this competition governs the effectiveness of the demon. 
In particular, we identify a parameter region where phonon-induced decoherence suppresses both the coherent transport contribution and the information flow toward the monitoring dot, suggesting that coherence can enhance the demon mechanism in this regime. By tracking information and transport properties across these crossovers, our model shows how coherent tunneling, decoherence, and incoherent phonon-assisted transport compete in an autonomous information engine, while clarifying which thermodynamic Lindblad description is appropriate in each regime.

\end{abstract}

\maketitle

\section{\label{sec:level1}Introduction\protect\\}

The thermodynamics of information has emerged as a central framework for understanding the role of information processing in nonequilibrium systems~\cite{parrondo2015thermodynamics}. 
This development has advanced along two fronts. On the one hand, extensive research has focused on non-autonomous setups~\cite{sagawa2009minimal,sagawa2008second,maruyama2009colloquium,cottet2017observing,naghiloo2018information}, where an external agent performs measurements and applies feedback, providing a controlled realization of Maxwell-demon-like operations. Experimental realizations~\cite{cottet2017observing,naghiloo2018information,koski2016maxwell} and measurement-driven classical-to-quantum transitions~\cite{annby2024maxwell} have been explored in this setting. On the other hand, information thermodynamics has been investigated in autonomous setups, where information flows continuously between interacting subsystems in the absence of external intervention~\cite{horowitz2014thermodynamics,yamamoto2016linear,mandal2012work,ehrich2023energy,leighton2025flow,barato2014stochastic,koski2015chip}.

Autonomous Maxwell's demons~\cite{strasberg2013thermodynamics,horowitz2014thermodynamics,mandal2012work,barato2014stochastic,ptaszynski2018autonomous} and Nonequilibrium demons~\cite{sanchez2019nonequilibrium,monsel2025autonomous}  have been studied and contrasted~\cite{freitas2021characterizing}, deepening our understanding of the thermodynamic role of information. 
The autonomous approach has been explored in molecular machines~\cite{leighton2025flow} and on mesoscopic quantum dot circuits~\cite{strasberg2013thermodynamics}. Such circuits can implement thermoelectric engines that convert heat currents into work against a chemical potential bias~\cite{thierschmann2015three}. In the two-quantum-dot proposal of Ref.~\cite{strasberg2013thermodynamics}, the device operates as an engine when one dot monitors the other, thereby acting as an autonomous Maxwell demon. This operation was formally interpreted within classical stochastic thermodynamics in Ref.~\cite{horowitz2014thermodynamics}, a framework that characterizes energy and information flows between subsystems and their relation to entropy production. An important extension of this approach was introduced in Ref.~\cite{ptaszynski2019thermodynamics}, where it was generalized to systems governed by Lindblad master equations.  
This extension enables the study of how genuinely quantum properties, such as coherence or entanglement, influence the information flows and the thermodynamic performance of the demon.

Despite these advances, less is known about how steady-state coherences in autonomous information engines are reflected in information flows, and how these flows change when the same device is driven toward an effectively classical transport regime. Here, we explore how information and thermodynamic properties change across two quantum-to-classical crossovers. By varying two independent control parameters, we track how coherent transport, information flow, and thermodynamic performance evolve as the device crosses over from a quantum-coherent to an effectively classical regime.

We consider a three-quantum-dot system in which one dot, coupled to a single reservoir, acts as a Maxwell demon by monitoring the occupation of a double dot that forms the transport channel between two reservoirs, enabling particles to flow against a chemical-potential bias.
The monitored system can be in nontrivial quantum states, with one electron in a superposition of localized states~\cite{friis2013fermionic,ptaszynski2023fermionic,dasenbrook2016single}. 
Within the Redfield master-equation framework, we investigate how quantum coherence in the double-dot influences the information flow toward the demon when the system is coupled to particle and phonon reservoirs.

Depending on the physical regime, characterized by the relative strength of coherent interdot tunneling and dissipative processes, the system either admits a description in terms of the standard secular Lindblad equation~\cite{breuer2002theory} or requires a more general approach compatible with the quantum features of the steady state. When the reservoir-induced decay rates are smaller than the coherent tunneling strength, the coherent oscillations are fast compared with the dissipative relaxation and can be averaged out under the secular approximation. In the opposite regime, where coherent tunneling is weak compared with the decay rates, some coherent oscillations are not resolved on the dissipative timescale, and a partial secular approximation is justified. Interestingly, in this regime the Redfield and partial secular Lindblad equations display steady-state coherences~\cite{trushechkin2021unified,potts2021thermodynamically}.

We explore two quantum-to-classical crossovers. The first is controlled by the coherent interdot tunneling strength and connects the partial- and full-secular regimes. In the full-secular limit, the steady state is fully described by populations obeying a classical master equation. The second crossover occurs within the partial-secular regime: we keep the coherent interdot tunneling weak and increase the effective coupling to a decohering phonon bath, which induces incoherent jumps between the quantum dots. In this case, decoherence drives the quantum-to-classical crossover.

A central aspect of our analysis is that the Redfield equation is used as a common benchmark to identify which Lindblad approximation provides the appropriate thermodynamic description in each regime. This allows us to distinguish changes in information flow arising from genuine physical mechanisms—coherent tunneling, decoherence, and phonon-assisted transport—from artifacts introduced by applying a secular approximation outside its domain of validity.

The paper is organized as follows. In Sec.~\ref{sec:levelmodel}, we introduce the three-quantum-dot system and its coupling to particle reservoirs and phonons. In Sec.~\ref{sec:masterEqs}, we discuss the effective open-system dynamics at the level of the Redfield equation and two of its Lindblad approximations: the \emph{global} master equation, obtained under a full secular approximation, and the \emph{semilocal} master equation, derived from a partial secular approximation. In Sec.~\ref{sec:NESSproperties}, we compare the steady states predicted by these three descriptions. In Sec.~\ref{sec:termo}, we review the thermodynamic framework, including the information flow between subsystems, as applied to the Lindblad descriptions. In Sec.~\ref{sec:results}, we show that the model operates as a three-terminal thermoelectric engine and, within a suitable parameter range, as an information engine. We analyze the two quantum-to-classical crossovers: In subsection \ref{sec:demoncoherent}, we increase the tunneling strength and investigate the demon as it crosses over from partial to full secular dynamics. In subsection \ref{sec:decoherent}, we do the analysis as we increase the coupling intensity to the phonon bath within the partial secular regime. 
We conclude in Sec.~\ref{sec:cocnlu}. Formal definitions of the notation and detailed derivations are referred to the Appendices.

\section{\label{sec:levelmodel}The Model\protect\\}

 \begin{figure}[h]
\includegraphics[scale = 0.50]{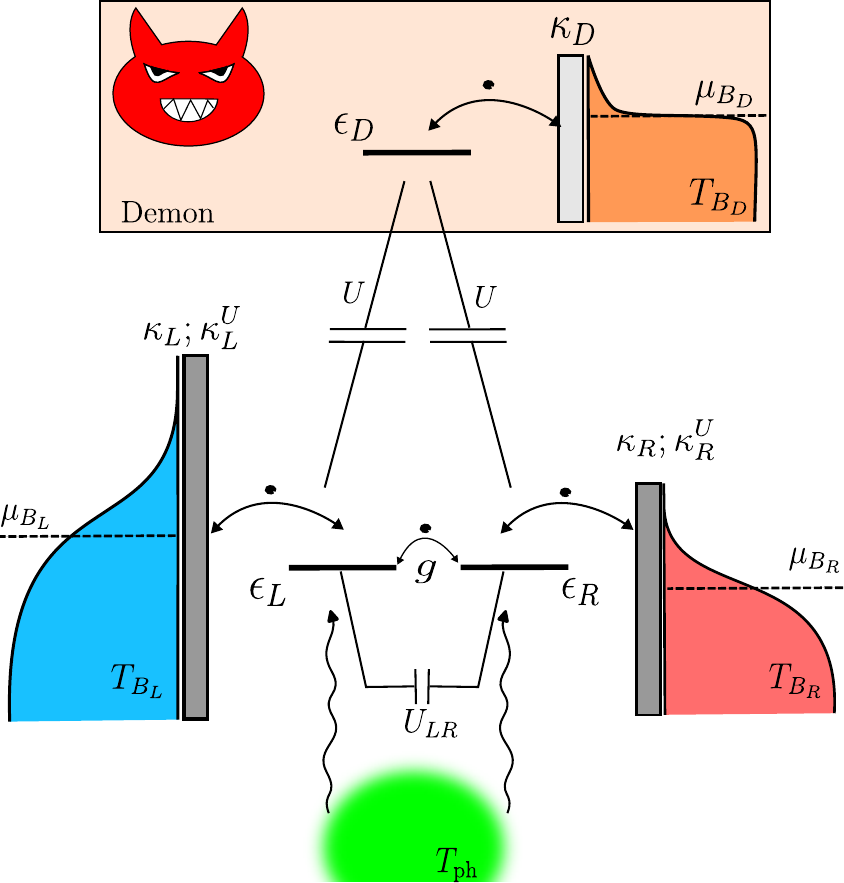}
\caption{\label{fig:model1} Scheme of the system composed of three quantum dots (thick black lines), coupled via Coulomb interactions $U_{LR}$ and $U$, indicated by capacitors, and via coherent tunneling ($g$) in the case of the left ($\epsilon_L$) and right ($\epsilon_R$) quantum dots. Each site is coupled to a different bath. The left and right dots constitute the working substance and are also coupled to the phonon bath (green). Furthermore, quantum dot $D$ ($\epsilon_D$), coupled to an independent bath (orange), acts as the Maxwell demon through its interaction with the working substance.}  
\end{figure}

We consider spinless electrons (or, equivalently, fully spin-polarized electrons), such that each quantum dot can be occupied by at most one electron.

Our system of interest $S$ consists of three quantum dots, labeled $L$, $R$, and $D$, as depicted in Fig.~\ref{fig:model1}. Dots $L$ and $R$ are coupled via coherent tunneling and interact through a Coulomb repulsion, while dot $D$ interacts capacitively with both $L$ and $R$. The Hamiltonian of the system is (we work in units where $\hbar = k_B = 1$)
\begin{align}
    \hat{H}_{S}  = & \, \epsilon(\hat{n}_{L} + \hat{n}_{R}) + \epsilon_{D}\hat{n}_{D} + g(\hat{d}^{\dagger}_{L}\hat{d}_{R} + \hat{d}^{\dagger}_{R}\hat{d}_{L}) \nonumber \\ 
    & + U_{LR}\hat{n}_{L}\hat{n}_{R} + U\hat{n}_{D}(\hat{n}_{L} + \hat{n}_{R}),
    \label{eq:HS}
\end{align}
where $\hat{n}_{i} = \hat{d}^{\dagger}_{i}\hat{d}_{i}$ ($i=L,R,D$) is the occupation operator. The fermionic operators satisfy the anticommutation relations $\{\hat{d}^{\dagger}_{i},\hat{d}_{j}\} = \delta_{ij}$. Dots $L$ and $R$ have equal onsite energies $\epsilon_L = \epsilon_R = \epsilon$, are tunnel-coupled with amplitude $g$, and interact via a Coulomb repulsion $U_{LR}$. Dot $D$, with onsite energy $\epsilon_D$, is Coulomb-coupled to both $L$ and $R$ with strength $U$. The eigenvalues and eigenstates of $\hat{H}_S$ are simply expressed in terms of the local bases in Appendix~\ref{miapp}. These two basis will play an important role in our discussions.

As we discuss below, the system can be viewed as bipartite: the $LR$ subsystem acts as the working substance, while the dot $D$ monitors the population in the double dot via the Coulomb interaction, effectively playing the role of a Maxwell demon.

The composite $LR$ system is weakly coupled to a phononic bath with Hamiltonian
\begin{equation*}
    \hat{H}_{\rm ph} = \sum_{k}\omega_{k}\hat{b}^{\dagger}_{k}\hat{b}_{k},
\end{equation*}
where the creation and annihilation operators satisfy the commutation relations $[\hat{b}_{k},\hat{b}^{\dagger}_{k'}]=\delta_{kk'}$. The interaction 
\begin{equation}
    \hat{V}_{\rm ph} = (\hat{d}^{\dagger}_{L}\hat{d}_{R} + \hat{d}^{\dagger}_{R}\hat{d}_{L}) \sum_{k}(h_{k}\hat{b}_{k} + h^{*}_{k}\hat{b}^{\dagger}_{k})=\hat{S}\hat{B}_{\mathrm{ph}},
    \label{eq:model_bathph}
\end{equation}
describes phonon-assisted tunneling between the sites $L$ and $R$ \cite{krause2011incomplete,rutten2009reaching}. This coupling conserves the system energy, i.e., $[\hat{H}_S,\hat{V}_{\rm ph}]=0$, and therefore induces jumps between states of equal energy.

Moreover, each quantum dot $i\in\{L,R,D\}$ is weakly coupled to an independent fermionic reservoir with Hamiltonian $\hat{H}_{B_i} = \sum_{\lambda}\epsilon_{i,\lambda}\hat{c}^{\dagger}_{i,\lambda}\hat{c}_{i,\lambda}$. The coupling describes electron tunneling between dot $i$ and bath $B_i$,
\begin{equation}
    \hat{V}_{i} = \sum_{\lambda}t_{i,\lambda}(\hat{d}^{\dagger}_{i}\hat{c}_{i,\lambda} + \hat{c}^{\dagger}_{i,\lambda}\hat{d}_{i} )=\hat{d}^{\dagger}_{i}\hat{B}_{i,+}+\hat{B}_{i,-}\hat{d}_{i},
    \label{eq:model_bath}
\end{equation}
where $\hat{B}_{i,+} = \sum_{\lambda} t_{i,\lambda} \hat{c}_{i,\lambda}$ and $\hat{B}_{i,-} = \sum_{\lambda} t_{i,\lambda} \hat{c}^{\dagger}_{i,\lambda}$.

In the following, we assume that the reservoirs are macroscopic and therefore possess a continuous spectrum. The open-system dynamics is determined by the system Hamiltonian and the bath spectral densities. For the phononic bath, we consider the spectral density
\[
J(\omega) = 2\pi \sum_k |h_k|^2\delta(\omega -\omega_k)
\]
to be Ohmic, i.e. $J(\omega) = J_{0}\omega e^{-\omega/\omega_c}$ with $\omega_{c}$ a cutoff frequency. Similarly, for the fermionic baths, we define the spectral density
\[
\kappa_{i}(\omega) = \sum_{\lambda}|t_{i,\lambda}|^{2} \delta(\omega-\epsilon_{i,\lambda}),
\]
which we take to have a Lorentzian form. 

We assume that all reservoirs are in thermal equilibrium. We denote the equilibrium state of reservoir $r$ as $\hat{\rho}_{r}^{\mathrm{eq}}$ ($r\in\{\mathrm{ph},B_D,B_L,B_R\}$). The phonon bath is characterized by a Bose-Einstein distribution at temperature $T_{\rm ph}$, while the reservoirs $B_i$ follow Fermi-Dirac distributions with chemical potentials $\mu_i$ and temperatures $T_i$. 

When these temperatures and chemical potentials differ, the system $S$ is driven to a nonequilibrium steady state at a characteristic relaxation rate. In this state, energy flows between all reservoirs, while particle transport occurs between $B_L$ and $B_R$.
This transport is enabled by the coherent tunneling term proportional to $g$ in the system Hamiltonian, as well as by phonon-assisted processes associated with the phononic bath. The coherent tunneling generates coherence and entanglement between the dots $L$ and $R$, while the phonon bath decoheres the state and suppresses quantum effects. Consequently, the quantum properties of the nonequilibrium steady state depend on the ratio of the phonon coupling strength, characterized by $J(\omega)$, to the tunneling strength $g$.

To analyze these properties of the nonequilibrium steady state, and thermodynamic properties like transport and information flows, we require an effective dynamical equation for the reduced density matrix of the three-quantum-dot system $S$. 

We conclude this section by specifying the parameter regime explored in this work. We consider equal temperatures for the phonon bath and the reservoirs coupled to dots $L$ and $R$, $T_{\mathrm{ph}}=T_{B_L}=T_{B_R}=T$, and drive the system out of equilibrium by imposing a chemical potential difference between $B_L$ and $B_R$, as well as a lower temperature $T_{B_D}$ for the reservoir coupled to the dot $D$.

 The system's relaxation rate is set by the bath couplings. Taking couplings to $B_L$, $B_R$, and $B_D$ of the same order, the characteristic relaxation rate is defined by $\kappa_L(\epsilon)\equiv \kappa_L$.

We assume a strong Coulomb repulsion $U_{LR}$, which restricts the $LR$ subsystem to single occupancy. Moreover, in order to operate in the regime where dot $D$ effectively monitors the $LR$ subsystem, we consider the Coulomb interaction energy $U$ to be large compared to the onsite energy, $U \gg \epsilon_D$.

We further consider the regime $|\epsilon_{D}|, U, U_{LR} \gg \kappa_L$. In this way, we can focus on two dynamical regimes determined by the ratio $g/\kappa_L$: the fast regime $g/\kappa_L \gg 1$, where fast coherent oscillations in the $LR$ subsystem justify the full secular approximation of the Redfield dynamics, and the slow regime $g/\kappa_L \lesssim 1$, where a partial secular approximation is required.

In our model, the system's Bohr frequencies are all different. However, in the regime $g/\kappa_L \lesssim 1$, incoherent transitions separated by an energy scale of order $g$ must be grouped together, as will be discussed below.

\section{Master equation description}
\label{sec:masterEqs}

We assume that the couplings between the system and the reservoirs are weak and that the reservoirs remain in equilibrium throughout the evolution. Under these Born--Markov conditions, the dynamics of $S$ can be approximated by a Redfield master equation that is local in time and second order in the system--reservoir couplings. 

Appendix~\ref{app:semilocal} contains the Redfield equation and its derivation for our system.
The dissipative terms are specified by a set of frequencies $\omega$, a set of decay rates $\gamma^{(r)}_l(\omega)$, and a set of jump operators $\hat A_{r,l}(\omega)$. 

We call the set of $\omega$ the relevant Bohr frequencies, and they play a very important role in our discussion. 
\begin{table}[h]
\centering
\renewcommand{\arraystretch}{1.4}
\begin{tabular}{l|c}
\toprule
 & \textbf{Relevant Bohr frequencies $\omega$} \\
\midrule
fermionic $B_D$ 
& $\epsilon_D,\;\; \epsilon_D + U,\;\; \epsilon_D + 2U$ \\
fermionic $B_L$ \& $B_R$ 
& $\epsilon_{\pm},\;\; \epsilon_{\pm} + U_{LR},\;\; 
   \epsilon_{\pm} + U,\;\; \epsilon_{\pm} + U + U_{LR}$ \\
bosonic ph 
& $0$ \\
\bottomrule
\end{tabular}
\caption{Relevant Bohr frequencies $\omega$ for fermionic $B_L,B_R$ and $B_D$ bath and for the bosonic phonon bath. Note that in row fermionic $B_L$ \& $B_R$, each symbol denotes a pair of frequencies because $\epsilon_{\pm}\equiv  \epsilon \pm g$. }
\label{tab:bohr-frequencies}
\end{table}
For our model, they are shown in Table \ref{tab:bohr-frequencies}.

 The $\gamma^{(r)}_l(\omega)$ rates are expressed in terms of the spectral densities $J(\omega)$ and $\kappa_i(\omega)$ of the reservoirs, and the Bose-Einstein distribution $\bar{n}_B(\omega)$ for the phonon bath and Fermi-Dirac distribution $f_i(\omega)$ for the fermionic reservoirs. They appear in the master equations evaluated at the frequencies of Table \ref{tab:bohr-frequencies}. Associated with the phonon bath, there is a single rate $\gamma_{\rm ph}(0)=\lim_{\omega\to 0}[1+\bar{n}_B(\omega)]J(\omega)=J_0T_{\rm ph}=J_0/\beta_{\mathrm{ph}}$,  a parameter that plays a central role in our discussion.
 The fermionic rates at these frequencies are specified in section~\ref{sec:results}.  

 The jumps operators $\hat A_{r,l}(\omega)$ are listed in Tables \ref{table:TablaII} and \ref{table:TablaIII} of Appendix~\ref{app:semilocal}.

The Redfield master equation captures the dissipative transitions induced by the reservoirs, but we seek a Lindblad-form master equation to formulate a consistent thermodynamic description of the nonequilibrium steady state~\cite{spohn1978entropy,spohn1978irreversible}.

\subsection{Global master equation (full secular approximation)}

When $|\epsilon_{D}|,U,U_{LR},g\gg \kappa_L$, all differences $|\omega-\omega'|$ between the Bohr frequencies in Table \ref{tab:bohr-frequencies} are greater than the characteristic relaxation rate $\kappa_L$. Under these conditions, the secular approximation to the Redfield equation is justified. This leads to 
a master equation in the GKLS form, 
\begin{equation*}
d_t\hat{\rho}_S
= -i[\hat H_S, \hat\rho_S]
+ \sum_r \mathcal{L}_r^{\mathrm{(f.s.)}}(\hat \rho_S),
\end{equation*}
where the dissipator associated with reservoir $r$ is given by 
\begin{equation}
\mathcal{L}_r^{\mathrm{(f.s.)}}(\hat \rho_S)
= \sum_{l,\omega} \gamma^{(r)}_l(\omega)\,
\mathcal{D}[\hat A_{r,l}(\omega)]\hat \rho_S,
\label{eq:dissip}
\end{equation}
with the standard superoperator $\mathcal{D}[\hat A]\hat \rho
= \hat A \hat \rho \hat A^\dagger
- \frac{1}{2}\{\hat A^\dagger \hat A,\hat \rho\}$.

 We call this equation the global master equation, since jump operators $\hat A_{r,l}(\omega)$ induce transitions between the (delocalized) eigenstates of $\hat H_S$ separated in energy by $\omega$. The global equation is explicitly written in~Appendix~\ref{app:redfieldglobal}.

\subsection{Semilocal master equation (partial secular approximation)}

When $g/\kappa_L \lesssim 1$, the secular approximation is no longer valid.
To obtain a master equation of GKLS form valid in the regime $g/\kappa_L\lesssim1$, one can resort to a partial secular approximation~\cite{potts2021thermodynamically,trushechkin2021unified}, in which near-degenerate frequencies (i.e., frequencies with $|\omega-\omega'|<\kappa_L$) are grouped into a single effective frequency and a single jump operator.  
This means that the four pairs of frequencies $\omega_{\pm}$ in the second line of Table \ref{tab:bohr-frequencies} are merged with the replacement $\epsilon_{\pm}\equiv  \epsilon \pm g\to \epsilon$. It also must be the case that $\gamma_l^{(r)}(\omega_+)\approx \gamma_l^{(r)}(\omega_-)$ for $r\in\{B_L,B_R\}$. The jump operator associated with the merged frequency is $\hat{A}_{r,l}(\omega_+)+\hat{A}_{r,l}(\omega_-)$.   
The resulting dissipators are denoted $\mathcal{L}_r^{\mathrm{(p.s.)}}$, and the collective jump operators associated with $B_L$ and $B_R$ are listed in Table~\ref{table:TablaIV} of Appendix \ref{app:appE}. They can be expressed in terms of operators acting on subsystems of $S$; the resulting master equation is thus commonly referred to as the \emph{semilocal Lindblad equation}~\cite{prech2023entanglement,potts2021thermodynamically}. Importantly, in our model, the semilocal Lindblad equation predicts steady-state coherences that coincide with those of the Redfield equation. Conversely, the steady state of the full secular master equation is diagonal in the energy basis.

\subsection{Thermodynamic consistency}
\begin{figure*}[t]
    \centering
    \includegraphics[scale = 0.98]{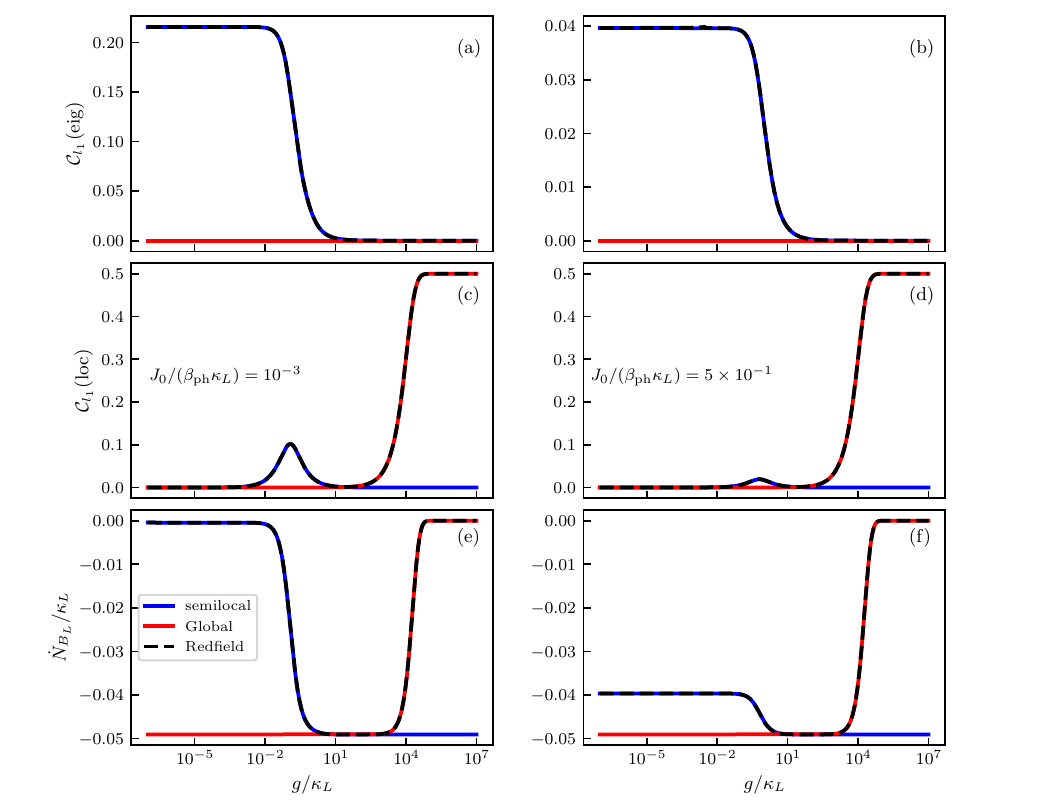}
    \caption{Coherence in the eigen basis $\mathcal{C}_{l_1}(\mathrm{eig})$, coherence in the local basis $\mathcal{C}_{l_1}(\mathrm{loc})$, and particle current, for two different values of phononic coupling $J_0/(\beta_{\mathrm{ph}}\kappa_L)= 10^{-3}$ and $J_0/(\beta_{\mathrm{ph}}\kappa_L)= 5 \times 10^{-1}$. Results are shown for the partial master equation (red), global master equation (blue), and Redfield equation(black dashed). $\mathcal{C}_{l_1}(\mathrm{eig})$ (a)-(b), $\mathcal{C}_{l_1}(\mathrm{loc})$ (c)-(d) and particle current $\dot{N}_{B_L}$(e)-(f) all as a function of $g/\kappa_L$. Parameters: $\kappa_L(\epsilon) =\kappa_R(\epsilon + U) = 1/100 $, $\kappa_L(\epsilon+U) =\kappa_R(\epsilon) = 1/600 $, $\kappa_{L}/\kappa_{R} = 6$, $\kappa_{D}/\kappa_{L} = 2$, $T/\kappa_{L} = 10000$, $T_{B_D}/\kappa_{D} = 100$, $\mu_L = -\mu_R$, $eV/T = 1$, $U_{LR}/T=5$, $U/T_{B_D}=20$, $\beta_{\mathrm{ph}}=\beta_{B_L} = \beta_{B_R}$, $\mu_{B_D} = 2$ and sites energies $\epsilon=0, \epsilon_D = \mu_{B_{D}}-U/2$ .}
    \label{fig:lindbladcomp1}
\end{figure*}

We have thus obtained two GKLS master equations valid in different parameter regimes.

Crucially, the fermionic rates satisfy the local detailed balance property $\gamma_+^{(B_j)}(\omega)/\gamma_-^{(B_j)}(\omega)=e^{-\beta_{j}(\omega-\mu_{j})}$. Thus, for each fermionic dissipator $\mathcal{L}_{B_j}^{\mathrm{(f.s.)}}$ or $\mathcal{L}_{B_j}^{\mathrm{(p.s.)}}$ we have
\begin{equation}
\mathcal{L}_{B_j}\!\left(
e^{-\beta_j(\hat H-\mu_j \hat N)}
\right) = 0,
\label{eq:DB}
\end{equation}
where $\beta_j$ and $\mu_j$ are the inverse temperature and chemical potential of reservoir $B_j$, and $\hat N=\hat{n}_L+\hat{n}_R+\hat{n}_D$ is the total particle-number operator, i.e., each dissipator admits the corresponding grand-canonical Gibbs state as a stationary state.

For the dissipators of the full secular Lindblad master equation $\mathcal{L}_{B_j}=\mathcal{L}_{B_j}^{\mathrm{(f.s.)}}$ we have $\hat H=\hat H_S$. Note that $[\hat{H}_S,\hat{N}]=0$. The thermodynamic consistency of this master equation is well established~\cite{Alicki79,spohn1978irreversible,breuer2002theory}. 

For the dissipators of the partial secular Lindblad master equation $\mathcal{L}_{B_j}=\mathcal{L}_{B_j}^{\mathrm{(p.s.)}}$, we have $\hat H=\hat H_{TD}$, which for our model is
\begin{equation*}
    \hat{H}_{TD} = \hat{H}_{S} - g (\hat{d}^{\dagger}_{L}\hat{d}_{R}+ \hat{d}^{\dagger}_{R}\hat{d}_{L}).
\end{equation*}
Note that $[\hat{H}_{TD},\hat{N}]=0$. The thermodynamic consistency of the semilocal master equation has been discuss more recently~\cite{potts2021thermodynamically}.
The reason for the appearance of $H_{TD}$ is that in the partial secular Lindblad equation, the jumps operators in $\mathcal{L}_{B_j}^{\mathrm{(p.s.)}}$ are eigenoperators of $H_{TD}$ and the induced energy changes correspond to the Bohr frequencies of $H_{TD}$. See Appendix \ref{app:appE}.

The energy changes induced by the jump operators are identified as the heat flow towards the system. In the partial secular Lindblad equation, these are coarse-grained at the energy scale ${\mathcal O}(g)$, and consequently, the predicted equilibrium state and heat current are coarse-grained as well. As discussed in \cite{potts2021thermodynamically,trushechkin2021unified}, this singles out $H_{TD}$, the so called thermodynamic Hamiltonian, as the appropriate operator to determine the internal energy, which keeps track of the energy changes associated with heat and work. The role of an operator $H_{TD}\neq H_S$ determined by  Eq.~\eqref{eq:DB} for thermodynamic quantities was also explored in ~\cite{lledo1,lledo2} in a related context.

In the nonequilibrium setup, when all baths are connected to the system, the jumps induced by the fermionic baths cannot resolve coherences at energy scales ${\mathcal O}(g)$. These coherences can therefore survive in the steady state, as captured by the semilocal dynamics.

The dissipator associated with the phonon bath $\mathcal{L}_{\rm ph}$ is the same in the full and partial secular approximation. It has a single jump operator $\hat{S}=d_L^\dagger d_R+d_R^\dagger d_L$ (see Appendix \ref{app:semilocal}1). Since $[\hat{S},\hat{H}_S]=[\hat{S},\hat{H}_{TD}]=[\hat{S},\hat{N}]=0$, it satisfies 
\begin{equation}
\mathcal{L}_{\rm ph}(e^{-\beta(\hat{H}-\mu\hat{N})})=0.
\label{eq:DB2}
\end{equation}

 Eqs.~\eqref{eq:DB} and \eqref{eq:DB2} ensure the thermodynamic consistency of the dynamics and provide a direct identification of heat and particle currents exchanged with each reservoir as we detail in section~\ref{sec:termo} after we discuss the properties of the nonequilibrium states.

\section{The nonequilibrium steady state: Redfield and its Lindblad approximations}
\label{sec:NESSproperties}

The Redfield equation is not of the GKLS form and, in general, may yield nonphysical density matrices because positivity is not guaranteed. We have numerically verified that, across all parameter regimes explored below, the Redfield steady states remain positive semidefinite and normalized to unit trace. We therefore use them as benchmarks to assess the validity of the corresponding Lindblad approximations.

We compare the steady states predicted by the Redfield, global, and semilocal master equations using the particle current between reservoirs $B_L$ and $B_R$, together with coherence measures in both the energy and local bases. The coherence measures are presented in Appendix~\ref{miapp}.

Figure~\ref{fig:lindbladcomp1} summarizes the comparison as a function of $g/\kappa_L$. A clear separation of regimes is observed. For $g/\kappa_L \lesssim 1$, the semilocal master equation is in excellent agreement with the Redfield solution, while the global master equation fails to reproduce both coherence and transport properties. In contrast, for $g/\kappa_L \gg 1$, the global master equation accurately captures the Redfield results, whereas the semilocal approximation breaks down. This crossover is consistent with the conditions derived in Sec.~\ref{sec:masterEqs}.

The origin of this behavior is reflected in the structure of the steady state. In the global (fully secular) description, the steady state is diagonal in the energy eigenbasis, and coherences are suppressed by construction. By contrast, the semilocal master equation retains nonsecular contributions that allow steady-state coherences.

To quantify these effects, we compute the $l_1$ norm of coherence in the energy basis,
\begin{equation*}
    \mathcal{C}_{l_1}(\mathrm{eig}) = |\tilde \alpha| + |\tilde \beta|,
\end{equation*}
as well as on the local basis,
\begin{equation*}
    \mathcal{C}_{l_1}(\mathrm{loc}) = |\alpha| + |\beta|.
\end{equation*}
Here, $\tilde\alpha,\tilde\beta$ are the nondiagonal elements of the steady-state density matrix in the eigenenergy basis and $\alpha,\beta$ in the local basis. These are the two nondiagonal terms allowed by the charge superselection rule. Importantly, $\alpha,\beta$ are purely imaginary.

Figures~\ref{fig:lindbladcomp1}(a)–(b) show the steady-state coherences in the energy eigenbasis. The semilocal and Redfield predictions are in good agreement over the full range of $g/\kappa_L$, whereas the global master equation suppresses these coherences and fails to reproduce the Redfield prediction for $g/\kappa_L \lesssim 1$. For large $g/\kappa_L$, all approaches predict vanishing coherences.

Figures~\ref{fig:lindbladcomp1}(c)–(d) show the coherences in the local basis. For $g/\kappa_L \lesssim 10^{1}$, the semilocal and Redfield predictions again coincide. The local maximum of $\mathcal{C}_{l_1}(\mathrm{loc})$ in the regime $g/\kappa_L \lesssim 1$ decreases as the phonon coupling increases, reflecting phonon-induced decoherence. For $g/\kappa_L \gtrsim 10^{1}$, the global master equation converges to the Redfield result, and the coherences approach an asymptotic value of $1/2$.

Figures~\ref{fig:lindbladcomp1}(e)–(f) show the particle current. For small $g/\kappa_L$, the semilocal equation reproduces the Redfield prediction, while the current increases with phonon coupling due to phonon-assisted jumps between $L$ and $R$. In the intermediate regime $10^{1} \lesssim g/\kappa_L \lesssim 10^{4}$, both Lindblad approximations provide a good description of the Redfield result. For $g/\kappa_L \gtrsim 10^{4}$, the global master equation captures the vanishing of the current, while the semilocal approximation fails to reproduce this behavior.

The suppression of transport at large $g$ can be understood from the energy spectrum: no system energy level lies within the transport window defined by the chemical potentials of $B_L$ and $B_R$. In this regime, the $LR$ subsystem approaches the antisymmetric state, consistent with the asymptotic value of the local coherence.

Overall, these results confirm that the semilocal master equation provides an accurate description in the low-$g/\kappa_L$ regime, capturing genuinely quantum features such as steady-state coherence, whereas the global master equation is valid in the opposite limit, where fast coherent dynamics justifies the secular approximation.

\section{Thermodynamics\label{sec:termo}} 

In this section, we apply the framework of nonequilibrium information thermodynamics~\cite{ptaszynski2019thermodynamics} to the Lindblad descriptions of our system,
\begin{equation}
d_t\hat \rho=-i[\hat H_S,\hat \rho]+\mathcal{L}(\hat \rho),
\label{lindbladThermo}
\end{equation}
where $\mathcal{L}=\sum_{r}\mathcal{L}_r$. We consider dissipators satisfying $\mathcal{L}_r\hat \rho_{r}^{\rm eq}=0$ where $\hat \rho_{r}^{\rm eq}=e^{-\beta_r(\hat{H}-\mu_r\hat{N})}/Z_r$ with $Z_r={\rm Tr}[e^{-\beta_r(\hat{H}-\mu_r\hat{N})}]$. Here $\hat{N}$ is the particle number operator and $\hat{H}$ is either the system Hamiltonian $\hat{H}_S$ or the thermodynamic Hamiltonian $\hat{H}_{TD}$, for $\mathcal{L}_r=\mathcal{L}_r^{\mathrm{(f.s)}}$ or $\mathcal{L}_r=\mathcal{L}_r^{\mathrm{(p.s)}}$, respectively. For the phonon bath, $\mu_\mathrm{ph}=0.$ 

Equation~\eqref{lindbladThermo} drives the system to a nonequilibrium steady state when the reservoirs have different temperatures and/or chemical potentials, leading to steady energy and particle flows.

Using Spohn's inequality~\cite{spohn1978entropy}, $-{\rm Tr}[\mathcal{L}_r(\hat \rho)(\ln\hat \rho-\ln\hat \rho_r^{\rm eq})]\geq 0$, and the time derivative of the von Neumann entropy $S=-{\rm Tr}[\hat \rho\ln\hat \rho]$, we can obtain a Clausius inequality
\[
d_t S-\sum_r \beta_r\dot Q_r=\dot\sigma\geq 0,
\]
where $\dot\sigma$ is the entropy-production rate, and the heat flow from reservoir $r$ is given by
$\dot Q_r={\rm Tr}[\mathcal{L}_r(\hat \rho)(\hat{H}-\mu_r\hat{N})]$~\cite{spohn1978entropy,spohn1978irreversible}.

Since $[\hat H,\hat H_S]=0$, the internal energy $E={\rm Tr}[\hat H\hat \rho]$ evolves as $d_t E=\sum_r \dot E_r$, with $\dot E_r={\rm Tr}[\hat H\mathcal{L}_r(\hat \rho)]$. Similarly, the particle number $N={\rm Tr}[\hat{N}\hat \rho]$ evolves as $d_t N=\sum_r \dot N_r$, with $\dot N_r={\rm Tr}[\hat{N}\mathcal{L}_r(\hat \rho)]$. This leads to the decomposition $\dot E_r=\dot W_r+\dot Q_r$, where $\dot W_r=\mu_r\dot N_r$ is the chemical work rate. Summing over all reservoirs yields the first law,
\[
d_tE=(\sum_r \dot W_r)+(\sum_r\dot Q_r)=\dot W+\dot Q.
\]

When we work with a semilocal equation, internal energy and heat are defined with the thermodynamic Hamiltonian. However, because the nonequilibrium steady state for our system satisfies ${\rm Tr}[\hat S\hat \rho]=2{\rm Re}(\alpha)+2{\rm Re}(\beta)=0$ (coherences are purely imaginary), we have ${\rm Tr}[\hat{H}_S\hat\rho]={\rm Tr}[\hat{H}_{TD}\hat\rho]$ and we can show that ${\rm Tr}[\hat{H}_S {\mathcal L}^{(p.s)}_r(\hat\rho)]={\rm Tr}[\hat{H}_{TD} {\mathcal L}^{(p.s)}_r(\hat\rho)]$. Therefore, heat and internal energy are given by the system Hamiltonian in our model.

We now consider a bipartition of the system into two subsystems labeled by ${\rm a}\in\{1,2\}$, where subsystem ${\rm a}=1$ corresponds to dot $D$ and subsystem ${\rm a}=2$ to the composite $LR$ system. To avoid confusion with the dot indices $i\in\{L,R,D\}$, we use roman indices ${\rm a}$ to denote subsystems.

The bath coupled to subsystem ${\rm a}=1$ is $B_D$, while the baths coupled to subsystem ${\rm a}=2$ are $\{\mathrm{ph},B_L,B_R\}$. Accordingly, we define
\[
\mathcal{L}_{\rm a} = \sum_{r \in {\rm a}} \mathcal{L}_r,
\]
where the sum runs over the baths connected to subsystem ${\rm a}$. The corresponding energy, heat, and work flows are defined as
\[
\dot E_{\rm a} = \sum_{r \in {\rm a}} \dot E_r, \quad
\dot Q_{\rm a} = \sum_{r \in {\rm a}} \dot Q_r, \quad
\dot W_{\rm a} = \sum_{r \in {\rm a}} \dot W_r.
\]
In particular, $\mathcal{L}_1 = \mathcal{L}_{B_D}$ and 
$\mathcal{L}_2 = \mathcal{L}_{\mathrm{ph}}+\mathcal{L}_{B_L}+\mathcal{L}_{B_R}$.

Following Ref.~\cite{ptaszynski2019thermodynamics}, the total entropy-production rate can be decomposed as $\dot\sigma=\dot\sigma_1+\dot\sigma_2\geq 0$, with
\[
\dot\sigma_{\rm a}=d_tS_{\rm a}-\beta_{\rm a}\dot Q_{\rm a}-\dot I_{\rm a}\geq 0,
\]
where $S_{\rm a}=-{\rm Tr}[\hat\rho_{\rm a}\ln\hat\rho_{\rm a}]$ is the entropy of subsystem ${\rm a}$, and $\beta_1=\beta_{B_D}$, $\beta_2=\beta=1/T$. The information flow is defined as
\begin{equation*}
    \dot{I}_{\rm a} = - {\rm Tr}( d_{t} \hat{\rho}_{\rm a} \ln \hat{\rho}_{\rm a}) + {\rm Tr}[ (\mathcal{L}_{\rm a}\hat{\rho})\ln \hat{\rho}],
\end{equation*}
and satisfies $d_t I=\dot I_1+\dot I_2$, where $I=S_1+S_2-S$ is the quantum mutual information~\cite{wilde2013quantum}.

We now apply these results to the steady state ($d_t=0$) and discuss the conditions under which the system operates as a three-terminal thermoelectric engine producing chemical work.

Since particles do not tunnel between dot $D$ and the $LR$ subsystem, $\dot N_{B_D}=0$, implying $\dot W_1=0$ and $\dot Q_1=\dot E_1=-\dot E_2$. The phonon bath exchanges no particles with the system, and the total particle current satisfies $\dot N_{B_L}+\dot N_{B_R}=0$. Therefore,
\[
\dot W=(\mu_L-\mu_R)\dot N_{B_L}.
\]
 If \(\mu_L>\mu_R\) and \(\dot N_{B_L}<0\), the system operates as a thermoelectric engine, with particles flowing from \(B_R\) to \(B_L\) against the electrochemical bias \(eV=\mu_L-\mu_R>0\). This behavior is compatible with the first law, \(\dot Q_{1}+\dot Q_2=-\dot W>0\), and the second law, \(\beta \dot Q_{2}+\beta_D\dot Q_1\leq 0\), provided that \(\dot Q_{2}>0\) and  \begin{equation}
  -\dot Q_{2}<\dot Q_{1}=\dot E_{1}<-\frac{T_{B_D}}{T}\dot Q_{2}<0.
  \label{ineqF}
  \end{equation}
  Thus, a flow against the bias ($\dot W<0$) is possible if heat flows through the system from the baths at temperature $T$ to the cold bath at $T_D$~\cite{callen1993thermodynamics,blundell2010concepts}.  

\subsection{Demon effect}

The local Clausius inequality
\begin{equation*}
\dot\sigma_{2}=-\beta_{2}\dot Q_{2}-\dot I_{2}\geq 0,
\end{equation*}
shows that a heat flow $\dot Q_2>0$ is allowed by an information flow $\dot I_2=-\dot I_1<0$.

 This identifies an information-engine mechanism, in which information extracted by subsystem ${\rm a}=1$ enables particle flow against a thermodynamic force.

Using $\dot Q_{2}=\dot E_{2}-\dot W$, this inequality can be written as
\[
T\dot\sigma_{2}=\dot W-\dot E_{2}-T\dot I_{2}\geq 0,
\]
which can be interpreted as a nonequilibrium free-energy inequality,
$\dot W\geq \dot{\mathcal{F}}_2$, with
$\dot{\mathcal{F}}_{\rm a}=\dot E_{\rm a}+T_{\rm a}\dot I_{\rm a}$.

At steady state, the condition $\dot W<0$ implies $\dot I_1>0$, although the converse is not necessarily true. More generally, while the thermoelectric-engine regime is characterized by $\dot W<0$, the operation is predominantly information-driven when $\dot I_1=-\dot I_2>0$ and the information contribution $T\dot I_1$, dominates over the energy exchanged between subsystems $|\dot E_1|$. In the limiting case where the energy exchange is negligible, $T\dot I_1>> |\dot E_1|$ this information-engine mechanism realizes the Maxwell-demon limit: dot $D$ acts as an autonomous Maxwell demon, and transport against the bias is enabled by information flow and negligible energy flow $\dot E_1$. Equation~\eqref{ineqF} further shows that achieving a small $\dot E_1$ at fixed $\dot Q_2$ requires $T_{B_D}\ll T$.

\section{\label{sec:results} Thermoelectric engine and information flows \protect\\}
\begin{figure}[h]
\includegraphics[scale = 1]{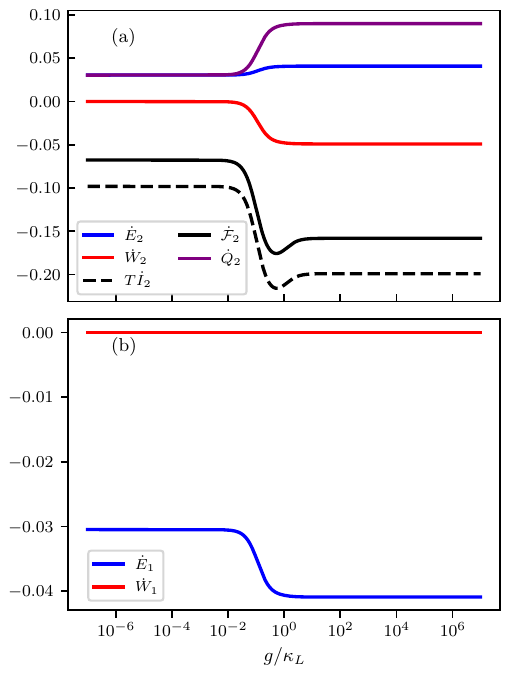}
\caption{\label{fig:termo}(a) Energy flow, Power, contribution of information, free energy rate, heat flow by the working substance. (b) Energy flow and power ($\dot{W}_{1}=0 \implies \dot{E}_{1}=\dot{Q}_{1}$), by the demon.}  
\end{figure}

In addition to a very small temperature ratio $T_{B_D}/T$, reducing $|\dot E_{\rm a}|$ requires an autonomous feedback mechanism 
that generates correlations between the demon $D$ and the working substance $LR$~\cite{strasberg2013thermodynamics,freitas2021characterizing,freitas2023information,kutvonen2016thermodynamics}. 
To this end, we introduce a frequency dependence in the spectral densities of baths $B_L$ and $B_R$ so that the corresponding tunneling rates into the $LR$ subsystem depend on the occupation of the demon. The strategy is the same for the global and semilocal Lindblad equations.  
Specifically, we assume a spectral density such that $\kappa_i(\epsilon^*) = \kappa_i(\epsilon^* + U +U_{LR}) = \kappa_i(\epsilon^* + U_{LR}) = \kappa_i$, $\kappa_i(\epsilon^* + U) = \kappa^{U}_i$, where $\epsilon^* = \epsilon \pm g$~\footnote{This can be achieved adding Lorentzian peaks centered at $\epsilon \pm g + U$.}. We further impose $\kappa_{L} = \kappa^{U}_{R} > \kappa_{R} = \kappa^{U}_{L}$.
In addition, the demon $D$ must rapidly detect the occupation of the working substance $LR$. This requires the demon to relax faster than the $LR$ subsystem, i.e, $\kappa_D(\omega)=\kappa_{D}> \kappa_R,\kappa_L$. The detection precision further relies on suppressing thermal fluctuations in the demon, which requires $\beta_{D} U \geq 1$. Together, these conditions enable the emergence of Maxwell-demon behavior in the three-quantum-dot system.

 

\subsection{\label{sec:demoncoherent}Quantum-to-classical transition driven by interdot tunneling strength \protect\\}

 In this section, we consider $J_{0}=0$, (i.e., no phonon bath) and analyze the performance of the demon as we increase $g/\kappa_L$ from the partial-secular regime (coherent Maxwell demon) to the full-secular regime (incoherent demon). 

In Fig.~\ref{fig:termo}(a), we plot the local free-energy rate of the working substance,
$\dot{\mathcal F}_{2}=\dot E_{2}+T\dot I_{2}$, together with its energetic and informational
contributions, as a function of $g/\kappa_L$. In the engine regime, the extracted chemical
power satisfies $\dot W<0$. The relevant thermodynamic bound is
$\dot W\geq \dot{\mathcal F}_{2}$, so work extraction requires the local free-energy rate
of the working substance to be negative.

As discussed in
Sec.~\ref{sec:termo}, compatibility with the first and second laws implies that, in the
engine regime, $\dot Q_1=\dot E_1<0$. Since the total energy is stationary, $\dot E_2=-\dot E_1>0$. Thus, the energetic
contribution to $\dot{\mathcal F}_{2}$ is positive. The negativity of
$\dot{\mathcal F}_{2}$ observed in Fig.~\ref{fig:termo}(a) is therefore entirely due to
the information term $T\dot I_2<0$.

Figure~\ref{fig:termo}(b) shows the corresponding thermodynamic quantities associated with
subsystem ${\rm a}=1$, namely the monitoring dot $D$. Since particles do not tunnel between
$D$ and the working substance, $\dot W_1=0$ and hence $\dot E_1=\dot Q_1$. In the parameter
range shown, this energetic exchange is small compared with the information contribution
driving the negative local free-energy rate in Fig.~\ref{fig:termo}(a). Thus, dot $D$ acts
as an autonomous Maxwell demon: it enables transport through information flow while exchanging a negligible amount of energy per unit time. A quantitative assessment of this condition over
the full parameter region is provided in Subsection~\ref{subsec:infodom} and Appendix~\ref{app:energy3d}.

\subsection{\label{sec:decoherent}Thermoelectric engine: phonon bath decoheres the working substance \protect\\}

\begin{figure}[h]
\includegraphics[scale = 1]{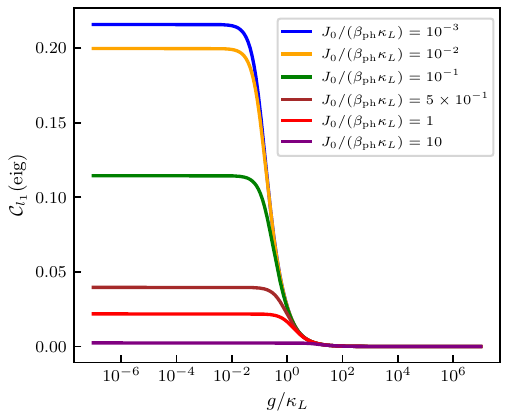}
\caption{\label{fig:phonon_cohe}Coherence $\mathcal{C}_{l_1}(\mathrm{eig})$ versus the hopping parameter $g/\kappa_L$, for different values of the phonon coupling strength $J_0/(\beta_{\mathrm{ph}} \kappa_{L})$.  Parameters: All parameters are the same of Fig.~\ref{fig:lindbladcomp1}, except for $J_0/(\beta_{\mathrm{ph}} \kappa_{L})$.} 
\end{figure}

The phonon-assisted tunneling allows particle current between the sites $L$ and $R$; this tunneling competes with coherent tunneling.
The dissipative effect also generates decoherence in the density matrix as
the Lindblad operator of the phononic bath generates dephasing~\cite{skinner1986pure,lidar2019lecture}. 
This can be seen in Fig.~\ref{fig:phonon_cohe}. As the phonon coupling $J_{0}/(\beta_{\mathrm{ph}} \kappa_L) $ increases, ${\mathcal C}_{l_1}(\rm{eig})$ decreases. 
Eventually, coherence is negligible for phonon interactions $J_{0}/(\beta_{\mathrm{ph}} \kappa_L)>10$. 

\begin{figure*}[t]
\includegraphics[scale = 1]{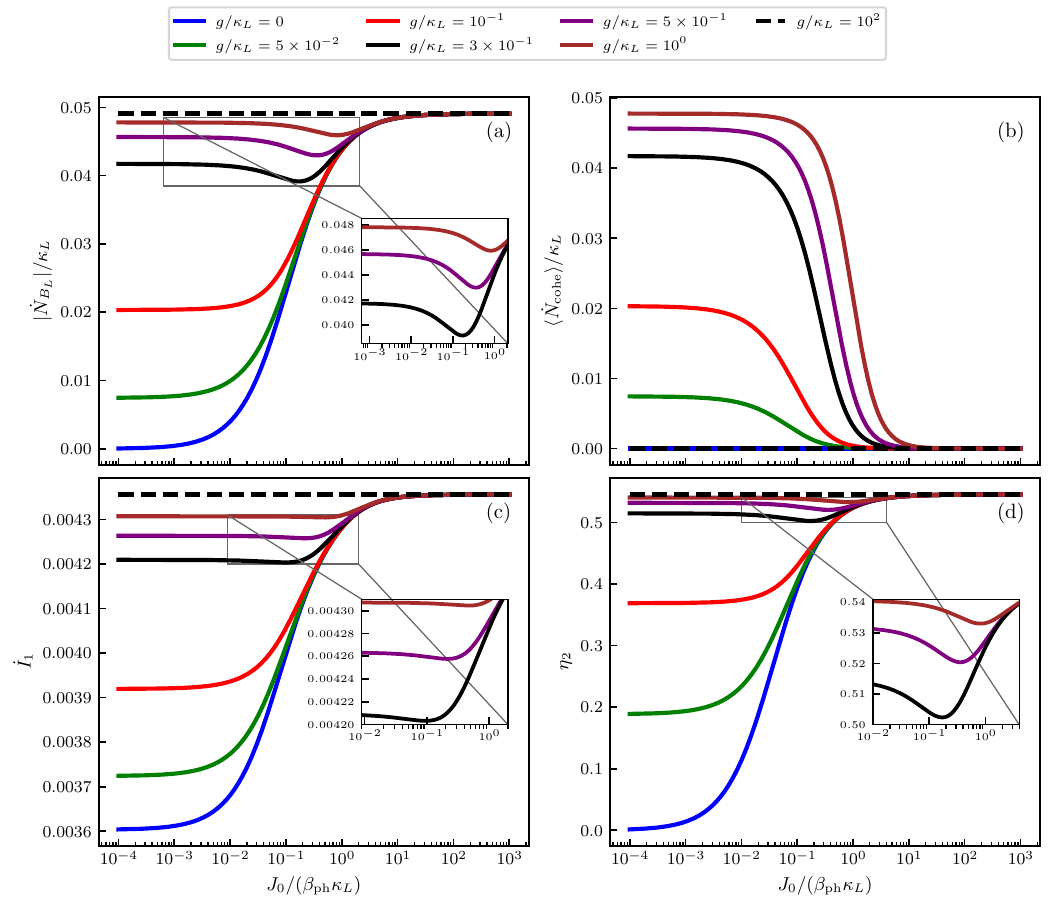}
\caption{\label{fig:phononeff}(a) Absolute value of the particle current $|\dot{N}_{B_L}|/\kappa_L$ versus phonon coupling $J_0/(\beta_{\mathrm{ph}} \kappa_L)$, for several parameters of coherent hopping $g/\kappa_L$. (b) $\langle \dot{N}_{\mathrm{cohe}} \rangle/\kappa_L$ against $J_0/(\beta_{\mathrm{ph}} \kappa_L)$. (c) Information flow received by the demon $\dot{I}_{1}$ versus phonon coupling $J_0/(\beta_{\mathrm{ph}} \kappa_L)$. (d) Efficiency of the working substance $\eta_2$ against $J_0/(\beta_{\mathrm{ph}} \kappa_L)$. Parameters: All parameters are the same of Fig.~\ref{fig:lindbladcomp1}, except for $g/\kappa_L$ and $J_0/(\beta_{\mathrm{ph}} \kappa_{L})$.}
\end{figure*}

The suppression of coherences induced by phonons allows us to identify a classical operation regime for $J_0/(\beta_{\mathrm{ph} }\kappa_L) \gtrsim 10^{0}$. This distinction enables a comparison between the quantum and classical-like behavior of the Maxwell demon. 

Figure~\ref{fig:phononeff}(a) shows the particle current, across the phonon-induced quantum-to-classical transition as a function of $J_{0}/(\beta_{\mathrm{ph}}\kappa_L)$ for several values of the coherent hopping $g$; for $J_0/(\beta_{\mathrm{ph}}\kappa_L) \gtrsim 10^{0}$, all curves converge to a common value, indicating that transport is dominated by phonon-assisted tunneling.
For most parameter sets, increasing the phonon coupling enhances the current. However, anomalous behavior appears for the purple, brown, and black dashed curves: In the regime $ 10^{-1} \lesssim J_0/(\beta_{\mathrm{ph}}\kappa_L) \lesssim 10^{0}$, the current is reduced due to decoherence as we show below. 

This behavior can be traced back to quantum coherence in the transport process. 
The coherent part of electron transport between the dots $L$ and $R$ can be obtained from a current operator
\begin{equation*}
    {\hat{J}}_{\mathrm{cohe}} = ig (\hat{d}^{\dagger}_L \hat{d}_R - \hat{d}^{\dagger}_R \hat{d}_L ),
\end{equation*}
derived from the continuity equation associated with particle conservation in the $LR$ system~\cite{wichterich2007modeling}.

Hence, we find the average coherent current as

\begin{equation*}
    \mathrm{Tr}({\hat{J}}_{\mathrm{cohe}} \hat{\rho}) = \langle \dot{N}_{\mathrm{cohe}}\rangle = 2g[\mathrm{Im}(\alpha) + \mathrm{Im}(\beta) ].
\end{equation*}

We obtained the right-hand side by replacing the nonequilibrium state obtained from the partial secular equation. If one uses the nonequilibrium state of the full secular equation, the trace is zero~\cite{spohn1978irreversible}.

As expected, comparing Figs.~\ref{fig:phononeff}(a) with \ref{fig:phononeff}(b), we observe $\dot N_{B_L}\to \langle \dot{N}_{\mathrm{cohe}}\rangle$ for $J_0/(\beta_{\mathrm{ph}}\kappa_L)\to 0.$
As shown in Fig.~\ref{fig:phononeff}(b), the curves exhibiting anomalous behavior correspond to those with the largest coherent particle currents. In the intermediate range
$10^{-1} \lesssim J_0/(\beta_{\mathrm{ph}}\kappa_L) \lesssim 10^{0}$ the coherent particle currents rapidly decrease.
For stronger phonon coupling, $J_0/(\beta_{\mathrm{ph}}\kappa_L) \gtrsim 10^{0}$, incoherent phonon-assisted tunneling dominates and the current increases again. See Fig.~\ref{fig:phononeff}(a).

Decoherence affects observable quantities beyond transport. Phonon-assisted tunneling also influences thermodynamic properties. A way to assess the performance of the device is through the information flow to the demon and the corresponding thermodynamic efficiency, defined as

\begin{equation*}
    \eta_2 = \left|\frac{\dot{W}_2}{\dot{Q}_2} \right|,
\end{equation*}
which measures the ratio between the work extracted from the subsystem $LR$ and the heat supplied to it.

Figure~\ref{fig:phononeff}(c) shows that the information flow to the demon generally increases with the phonon coupling for most values of $g/\kappa_L$, reflecting the enhancement of particle transport by phonon-assisted processes. However, for the purple, black, and brown curves, $\dot I_1$ decreases in the intermediate regime
$10^{-1} \lesssim J_0/(\beta_{\mathrm{ph}}\kappa_L) \lesssim 10^{0}$,
where phonon-induced decoherence suppresses the coherent contribution to transport and information flows. 

The corresponding efficiency, shown in Fig.~\ref{fig:phononeff}(d), follows the same competition. Although stronger phonon coupling generally increases $\eta_2$ by enhancing incoherent transport, the same intermediate-coupling regime shows a reduction in efficiency for the curves with the largest coherent contribution. This behavior reflects the detrimental effect of decoherence before phonon-assisted transport becomes dominant.

Overall, Fig.~\ref{fig:phononeff} demonstrates that thermodynamic performance is influenced by quantum coherences. While phonon-assisted processes enhance transport and efficiency when $g/\kappa_L$ is negligible, the associated phonon-induced decoherence suppresses the quantum contributions existing for intermediate $g/\kappa_L$ and reduces performance.  

\subsection{Information-dominated engine regime}
\label{subsec:infodom}

The previous results show representative cuts where $\dot N_{B_L}<0$, corresponding to transport against the applied bias. This behavior persists over the two-dimensional parameter region explored here, as shown in Appendix~\ref{app:energy3d}. We therefore focus on the quantity that characterizes the information-driven engine, namely, the ratio
\[
\left|\frac{\dot E_1}{T\dot I_1}\right|.
\]

\begin{figure}[h]
\includegraphics[scale = 1]{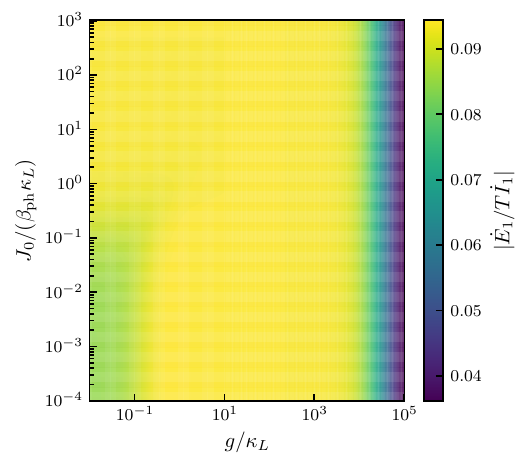}
\caption{\label{fig:ratio}  Plot of the ratio $|\dot{E}_1/T\dot{I}_1|$ as a function of $g/\kappa_L$ and $J_0/(\beta_{\mathrm{ph}} \kappa_L)$. From $g/\kappa_L \lesssim 10^{1}$, data were obtained with the semilocal equation. From $g/\kappa_L \gtrsim 10^{1}$, data were obtained with the global equation. Parameters: All parameters are the same of Fig.~\ref{fig:lindbladcomp1}, except for $g/ \kappa_{L}$ and $J_0/(\beta_{\mathrm{ph}} \kappa_L)$.}
\end{figure}

Figure~\ref{fig:ratio} shows that this ratio remains much smaller than unity throughout the relevant parameter region.  This supports the interpretation of dot $D$ as an autonomous Maxwell demon and of the full device as a thermoelectric information engine operating in the Maxwell-demon limit.

\section{\label{sec:cocnlu}Conclusions}

We have investigated a three-terminal thermoelectric engine based on a system of three quantum dots, which can be naturally described as a bipartite setup. In this picture, the quantum dot $D$  monitors the occupation of the $LR$ subsystem via Coulomb interaction. 

We adopted the Redfield equation as a consistent description of the system dynamics across the parameter regimes considered. In the regime of large tunneling $g$, the secular approximation faithfully reproduces the Redfield dynamics, and the steady state is effectively classical, with no role played by coherence. Accordingly, the information flow is well captured by classical stochastic thermodynamics. In contrast, for small $g$, a partial secular approximation is required to correctly reproduce the Redfield steady state. This approximation retains a Lindblad structure, in which the steady state of the reduced $LR$ system preserves coherences. 

Consequently, while the dot monitors a classical working substance for large $g$, it probes a quantum one in the small-$g$ regime. Our analysis of steady-state information flows reveals, for low bias and in a range of $g$ covering the classical (diagonal) and quantum states of the $LR$ subsystem, an engine regime characterized by particle transport against the chemical-potential bias between dots $R$ and $L$. This regime exhibits a positive information flow $\dot I_1$ toward $D$ and negligible energy flow ($\dot E_1\ll T\dot I_1$), which establishes a clear interpretation of the device as an autonomous information engine, in which transport emerges from the combined effect of the information flow toward $D$ and the modulation of reservoir transition rates. 

We have also examined the role of an additional phonon bath coupled to the $LR$ subsystem with strength $J_0$. For large $g$, the phonons do not have an effect because there are no transitions between nondegenerate states. However, for small $g$, the bath induces both incoherent transport and decoherence in the $LR$ subsystem. If $g\ll \kappa_L$, coherences in the $LR$ system are weak, decoherence is ineffective, and thus increasing $J_0$ enhances transport via phonon-assisted jumps. If $g\lesssim \kappa_L$, coherences in the $LR$ subsystem are important; here, as $J_0$ grows, decoherence initially suppresses coherent transport, while at larger coupling strengths, incoherent processes dominate, effectively driving a quantum-to-classical transition. Notably, the interpretation of the device as an autonomous information engine remains valid across both regimes. However, the underlying physical mechanisms change qualitatively along this crossover: for small $J_0$, transport contains a sizable coherent-current contribution correlated with the information flow toward $D$, whereas for large $J_0$, coherences are suppressed and transport is dominated by incoherent phonon-assisted processes.

Information flows are affected similarly. If $g\ll \kappa_L$, increasing $J_0$ enhances the information flow, whereas if $g\lesssim \kappa_L$, decoherence initially reduces it, thereby affecting the demon's ability to monitor the $LR$ subsystem. Interestingly, for sufficiently large phonon coupling, the phonon bath enhances the total information flow, reflecting the growing importance of incoherent transport mechanisms.

Overall, our results provide a unified picture linking thermoelectric performance, information flow, and the quantum-to-classical crossover in autonomous nanoscale engines, and clarify the conditions under which a quantum system can be meaningfully interpreted as a Maxwell's demon.

\begin{acknowledgments}
F. B. Thanks, Fondecyt project 1231210. M. B.
S. and J.M acknowledge partial support from Fondecyt project 1231210.

\end{acknowledgments}

\appendix

\section{Hamiltonian, eigenstates, notation}
\label{miapp}
The eigenenergies and eigenstates of the Hamiltonian in Eq.\eqref{eq:HS} appear in Table \ref{miTabla} with the total (eigen) particle number.
\begin{table}[t]
\centering
\scriptsize
\setlength{\tabcolsep}{3pt}
\renewcommand{\arraystretch}{0.9}
\begin{tabular}{cccccc}
\hline
$\ket{0,0}$ 
& $\ket{0,1}$ 
& $\ket{\pm,0}$ 
& $\ket{2,0}$ 
& $\ket{\pm,1}$ 
& $\ket{2,1}$ \\

$0$ 
& $\epsilon_D$ 
& $\epsilon_{\pm}$ 
& $2\epsilon+U_{LR}$ 
& $\epsilon_{\pm}+\epsilon_D+U$ 
& $2\epsilon+U_{LR}+\epsilon_D+2U$ \\

$0$ 
& $1$ 
& $1$ 
& $2$ 
& $2$ 
& $3$ \\
\hline
\end{tabular}
\caption{First row: eigenstates, Second row: eigenenergies. Third row: total number.}
\label{miTabla}
\end{table}
These eigenstates are product states of the state of the $LR$ state, first element in the ket, and the occupation of system $D$ (number state $\ket{D},\, D\in (0,1)$), the second element in the ket.
In terms of the particle number states $\ket{L,R},\, L,R\in(0,1)$ of $\hat n_L+\hat n_R$ we have: 
\begin{align}
\ket{0}&=\ket{0,0}\nonumber,\\
\ket{\pm}&=(\ket{0,1}\pm\ket{1,0})/\sqrt{2}\nonumber,\\
\ket{2}&=\ket{1,1}.
\end{align}

The charge superselection rule~\cite{wick1952intrinsic} allows the following non-vanishing off-diagonal density matrix elements 
$\tilde \alpha = \langle +,0|\hat{\rho}|-,0\rangle$ and $\tilde \beta = \langle +,1|\hat{\rho}|-,1\rangle$ in the eigenenergy basis. 
It is also interesting to consider the non vanishing coherences $\alpha = \langle 1,0,0|\hat{\rho}|0,1,0\rangle$ and $\beta = \langle 1,0,1|\hat{\rho}|0,1,1\rangle$ in the local basis $|L,R,D\rangle$.

\section{\label{app:semilocal}Redfield equation for the three quantum dot system}

Starting from the system--bath couplings in Eqs \eqref{eq:model_bathph} and \eqref{eq:model_bath}, which we write 
\begin{equation*}
\hat V_r=\sum_l \hat A_{r,l}\hat B_{r,l},
\end{equation*}
the Born--Markov approximation leads to the Redfield equation for the reduced density matrix of the system in the interaction picture,
\begin{widetext}
\begin{equation}
\partial_t\hat{\rho}_S(t)=
\sum_r\sum_{l,\omega,\omega'} 
e^{i(\omega'-\omega)t}\Gamma_l^{(r)}(\omega)
\Big(
\hat{A}_{r,l}(\omega)\hat{\rho}_S(t)\hat{A}_{r,l}^\dagger(\omega')
-
\hat{A}_{r,l}^\dagger(\omega')\hat{A}_{r,l}(\omega)\hat{\rho}_S(t)
\Big)
+\mathrm{H.c.},
\label{eq:REdfield}
\end{equation}
\end{widetext}
where the coefficients
\begin{equation}
\label{eq:gamma}
\Gamma^{(r)}_l(\omega)
= \int_0^\infty dt\, e^{i \omega t}
\mathrm{Tr}\!\left[
e^{i\hat H_rt}\hat B_{r,l}^\dagger e^{-i\hat H_rt} \hat B_{r,l}\hat \rho_{r}^{\mathrm{eq}}
\right],
\end{equation}
are bath correlation functions. The operators $\hat A_{r,l}(\omega)$ are the components of the system operators in the Bohr-frequency decomposition,
\begin{equation}
\label{eq:C4}
e^{i\hat H_St}\hat A_{r,l}e^{-i\hat H_St}
=\sum_\omega e^{i\omega t}\hat A_{r,l}(\omega).
\end{equation}
The set of frequencies on the right-hand side is the set of relevant Bohr frequencies, see Ref.~\cite{breuer2002theory}.

Transforming back to the Schrodinger picture, the Redfield equation is given by
\[
\partial_t\hat \rho_S=-i[\hat H,\hat \rho_S]+\sum_r \tilde{\mathcal{L}}_{r}\hat \rho_S,
\]
where $\hat H=\hat H_S+\delta \hat H_{LS}$ is the sum of the system Hamiltonian plus a Lamb shift term $\delta \hat H_{LS}$, that depends on the values $\mathrm{Im}[\Gamma^{(r)}_l(\omega)]$ and 
\begin{align}
    \tilde{\mathcal{L}}_{r}\hat \rho_S(t) = & \sum_{l,\omega,\omega'} \frac{(\gamma^{(r)}_{l}(\omega) + \gamma^{(r)}_{l}(\omega'))}{2} \hat{A}_{r,l}(\omega) \hat \rho_{S}(t)\hat{A}^{\dagger}_{r,l} (\omega')  \nonumber\\
     & -\sum_{l,\omega,\omega'} \frac{\gamma^{(r)}_{l}(\omega)}{2}\hat{A}^{\dagger}_{r,l} (\omega')\hat{A}_{r,l}(\omega)\hat \rho_{S}(t) \nonumber\\
     & + \sum_{l,\omega,\omega'}\frac{\gamma^{(r)}_{l}(\omega')}{2}\hat \rho_{S}(t)\hat{A}^{\dagger}_{r,l} (\omega')\hat{A}_{r,l}(\omega), 
     \label{eq:dissipRedfield}
\end{align}
with $\gamma^{(r)}_{l}(\omega) = \mathrm{Re}[\Gamma^{(r)}_{l}(\omega)]$. In our simulations, the Lamb-shift term $\delta \hat H_{LS}$ is neglected.

This expression is not in the GKLS form, so it does not ensure positivity and trace preservation. Nevertheless, once we obtain the jump operators $\hat{A}_{r,l}(\omega)$, we simulate the Redfield equation, including the contribution from each bath, and find positive states in the parameter range we explored.

\subsection{Explicit expressions for the phonon bath}

The phonon-bath system coupling Eq.~\eqref{eq:model_bathph} features the single bath operator $\hat{B}_{\mathrm{ph}}=\sum_k(h_k\hat{b}_k+h_k^*\hat{b}_k^\dagger)$,  which couples to the system operator  $\hat{S}=\hat{d}_R^\dagger \hat{d}_L+\hat{d}_L^\dagger \hat{d}_R$. This system operator satisfies $\hat{S} = e^{i\hat{H}_{S}t}\hat{S}e^{-i\hat{H}_S t}$, therefore, we find that the only relevant Bohr frequency is $\omega=0$, see Eq.~\ref{eq:C4}, and that the associated jump operator is $\hat{A}_{\mathrm{ph}}(\omega=0)=\hat{S}$, which acts non-trivially in the single particle space of the subsystem $LR$. Thus, the phononic bath induces jumps between dots $L$ and $R$ or vice versa, independently of the population of $D$.

Replacing $\hat{B}_{\mathrm{ph}}$ in Eq.\eqref{eq:gamma} and taking the real part, we obtain the phononic decay rat e

\begin{equation*}
    \gamma_{\mathrm{ph}}(\omega) = J(\omega)[1 + n_{B}(\omega)].
\end{equation*}

Here $J(\omega) = 2\pi \sum_k |h_k|^{2}\delta(\omega - \omega_k)$ denotes the spectral density and $n_B(\omega) = (\mathrm{exp}(\beta_{\mathrm{ph}}\omega )  -1)^{-1}$ is the Bose-Einstein distribution. In this work we consider an Ohmic spectral density $J(\omega) = J_0 \omega e^{-\omega/\omega_c}$ with $\omega_c$ a cutoff frequency \cite{barr2024spectral,schaller2014open,mascherpa2017open}. In the limit $\omega \to 0$ this leads to $\gamma_{\mathrm{ph}}(\omega \to 0) = J_0/\beta_{\mathrm{ph}}$.

Because there is a single relevant Bohr frequency $\omega=0$ for the phonon bath, the dissipator Eq.\eqref{eq:dissipRedfield} gives
\begin{equation*}
    \mathcal{L}_{\mathrm{ph}}(\hat{\rho}) = \frac{J_0}{\beta_{\mathrm{ph}}} \left[ \hat{S} \hat{\rho}\hat{S}^{\dagger} - \frac{1}{2} \{ \hat{S}^{\dagger}\hat{S},\hat{\rho} \}  \right],
\end{equation*}
which is already in GKLS form.

\subsection{Explicit expressions for the fermionic baths}

In the expression for the coupling with the fermionic reservoir $B_j$ (with $j\in \{L,R,D\}$), Eq.~\eqref{eq:model_bath}, the corresponding operators $\hat B_{j,l=\pm}$ are $\hat{B}_{j,+}=\sum_\lambda t_{j,\lambda}\hat{c}_\lambda$ and $\hat{B}_{j,-}=\sum_\lambda t_{j,\lambda}\hat{c}^\dagger_\lambda$. 

Replacing $\hat{B}_{j\pm}$ in Eq.\eqref{eq:gamma} and taking the real part we obtain the decay rates
\begin{equation}
    \gamma^{(B_j)}_{+}(\epsilon) = \kappa_{j}(\epsilon)f_{j}(\epsilon), \hspace{5mm}   \gamma^{(B_j)}_{-}(\epsilon) = \kappa_{j}(\epsilon)(1-f_{j}(\epsilon)),
    \label{eq:fermirates}
\end{equation}
with the tunneling rates $\kappa_j(\omega) = \sum_{\lambda}t^{2}_{j,\lambda} \delta(\omega - \epsilon_{j,\lambda})$ and the Fermi-Dirac distribution $f_j(\omega)=(1+\mathrm{exp}(\beta_j (\omega - \mu_j) ))^{-1}$.

The operators $\hat B_{j,l=\pm}$ couple to the system via the operators $d_j^\dagger$ and $d_j$ in Eq.~\eqref{eq:model_bath}. The expansion \eqref{eq:C4} for these operators determine the jumps operators $\hat{A}_{r,l}(\omega)$. They are given in Table \ref{table:TablaII} for $r=B_D$ and Table \ref{table:TablaIII} for $r=B_L,B_R$. We note that $\hat A_{r,-}(\omega)=\hat A_{r,+}^\dagger(-\omega)$. 

With Eq.\eqref{eq:fermirates} and Tables \ref{table:TablaII} and \ref{table:TablaIII} we obtain the Redfield dissipators \eqref{eq:dissipRedfield} for $r=B_L,B_R,B_D$. 

We now perform the analytical steps to obtain Tables \ref{table:TablaII} and \ref{table:TablaIII}.

We introduce the fermionic operators
\begin{equation}
    \hat{d}_{-} = \frac{1}{\sqrt{2}}(\hat{d}_R - \hat{d}_L ), \hspace{5mm} \hat{d}_{+} = \frac{1}{\sqrt{2}}(\hat{d}_R + \hat{d}_L ), \label{app:cambiobase}
\end{equation}
to diagonalize the three–quantum-dot Hamiltonian and calculate\eqref{eq:C4} for the fermionic operators. In terms of Eq.\eqref{app:cambiobase}, Eq.~\eqref{eq:HS} becomes
\begin{equation*}
    \hat{H}_{S} = \epsilon_{+}\hat{n}_+ + \epsilon_{-}\hat{n}_{-} +\epsilon_{D}\hat{n}_{D} + U(\hat{n}_- + \hat{n}_+ )\hat{n}_{D} + U_{LR}\hat{n}_{+}\hat{n}_{-},  
\end{equation*}
where $\hat{n}_{\pm}=\hat{d}_{
\pm}^\dagger \hat{d}_{\pm}$ and $\epsilon_{\pm} = \epsilon\pm g$. 

We first compute Eq.\eqref{eq:C4} for $\hat{d}_D$

\begin{align}
    & e^{i\hat{H}_S t} \hat{d}_{D}e^{-i\hat{H}_S t}  = e^{-i\epsilon_D t} \hat{d}_D (1-\hat{n}_{+})(1-\hat{n}_{-}) \nonumber\\ &+ e^{-i(\epsilon_D + U)t}\hat{d}_D [ (1-\hat{n}_{+})\hat{n}_{-} + \hat{n}_{+}(1-\hat{n}_{-}) ] \nonumber\\ 
     & + e^{-i(\epsilon_D + 2U)t}\hat{d}_{D}\hat{n}_{+}\hat{n}_{-}.
     \label{eq:jumpDemon}
\end{align}

The relevant Bohr frequencies and the jump operators displayed in Eq.~\eqref{eq:jumpDemon} are
summarized in Table \ref{table:TablaII}.

\begin{table}[h]
\centering
\renewcommand{\arraystretch}{1.4}
\begin{tabular}{|c|c|}
\hline
\textbf{Frequency} & $ D$ \\
\hline
$\epsilon_D$ 
& $\hat d_D(1-\hat n_{+})(1-\hat n_{-})$ \\
\hline
$\epsilon_D + U$ 
& $\hat d_D\!\left[(1-\hat n_{+})\hat n_{-} + \hat n_{+}(1-\hat n_{-})\right]$ \\
\hline
$\epsilon_D + 2U$ 
& $\hat d_D\,\hat n_{+}\hat n_{-}$ \\
\hline
\end{tabular}
\caption{Bohr frequencies associated with the reservoir coupled to dot $D$ and their corresponding jump operators.}
\label{table:TablaII}
\end{table}

To obtain Eq.\eqref{eq:C4} for $\hat d_L$ and $\hat d_R$ we first compute

\begin{equation}
\begin{aligned}
    & e^{i\hat{H}_S t} \hat{d}_{\pm}e^{-i\hat{H}_S t}  = e^{-i\epsilon_\pm t} \hat{d}_\pm (1-\hat{n}_{D})(1-\hat{n}_{\mp})\\
    & + e^{-i(\epsilon_\pm + U_{LR})t}\hat{d}_\pm (1-\hat{n}_{D})\hat{n}_{\mp} \\
    & + e^{-i(\epsilon_\pm + U)t}\hat{d}_\pm (1-\hat{n}_{\mp})\hat{n}_{D} + e^{-i(\epsilon_\pm + U + U_{LR})t}\hat{d}_{\pm}\hat{n}_{D}\hat{n}_{\mp}.
\end{aligned}
\label{app:jump_op_global0}
\end{equation}

From Eq.\eqref{app:cambiobase} and \eqref{app:jump_op_global0} we obtain
\begin{align}
    & e^{i\hat{H}_S t} \hat{d}_{L}e^{-i\hat{H}_S t}  = \frac{1}{\sqrt{2}} \sum_{\alpha,\beta=\pm;\alpha\neq \beta}\alpha \left( e^{-i\epsilon_\alpha t} \hat{d}_\alpha (1-\hat{n}_{D})(1-\hat{n}_{\beta}) \right.  \nonumber \\
    & \left. + e^{-i(\epsilon_\alpha + U_{LR})t}\hat{d}_\alpha (1-\hat{n}_{D})\hat{n}_{\beta} \right.  \nonumber \\
    &\left. + e^{-i(\epsilon_\alpha + U)t}\hat{d}_\alpha (1-\hat{n}_{\beta})\hat{n}_{D} + e^{-i(\epsilon_\alpha + U + U_{LR})t}\hat{d}_{\alpha}\hat{n}_{D}\hat{n}_{\beta}\right),
    \label{app:jump_op_global1}
\end{align}
and
\begin{align}
    & e^{i\hat{H}_S t} \hat{d}_{R}e^{-i\hat{H}_S t}  = \frac{1}{\sqrt{2}} \sum_{\alpha,\beta=\pm;\alpha\neq \beta} \left( e^{-i\epsilon_\alpha t} \hat{d}_\alpha (1-\hat{n}_{D})(1-\hat{n}_{\beta}) \right. \nonumber \\
    & \left. + e^{-i(\epsilon_\alpha + U_{LR})t}\hat{d}_\alpha (1-\hat{n}_{D})\hat{n}_\beta \right. \nonumber \\
    &\left. + e^{-i(\epsilon_\alpha + U)t}\hat{d}_\alpha (1-\hat{n}_{\beta})\hat{n}_{D} + e^{-i(\epsilon_\alpha + U + U_{LR})t}\hat{d}_{\alpha}\hat{n}_{D}\hat{n}_{\beta}\right).
    \label{app:jump_op_global2}
\end{align}

The relevant Bohr frequencies and the jump operators displayed in Eq.~\eqref{app:jump_op_global1} and \eqref{app:jump_op_global2} are
summarized in Table \ref{table:TablaIII}

\begin{table}[h]
\centering
\begin{tabular}{|c|c|c|}
\hline
\textbf{Frequency} & \textbf{L} & \textbf{R} \\
\hline
$\epsilon_{+}$ & $\hat{d}_+ (1-\hat{n}_{D})(1-\hat{n}_{-})$ & $\hat{d}_+ (1-\hat{n}_{D})(1-\hat{n}_{-})$ \\
\hline
$\epsilon_{-}$ & $-\hat{d}_- (1-\hat{n}_{D})(1-\hat{n}_{+})$ & $\hat{d}_- (1-\hat{n}_{D})(1-\hat{n}_{+})$  \\
\hline
$\epsilon_{+}+U$ & $\hat{d}_+ (1-\hat{n}_{-})\hat{n}_D$ &  $\hat{d}_+ (1-\hat{n}_{-})\hat{n}_D$ \\
\hline
$\epsilon_{-}+U$ & $-\hat{d}_- (1-\hat{n}_{+})\hat{n}_D$ & $\hat{d}_- (1-\hat{n}_{+})\hat{n}_D$ \\
\hline
$\epsilon_{+}+U_{LR} $ & $\hat{d}_+ \hat{n}_{-}(1-\hat{n}_D)$ & $\hat{d}_+ \hat{n}_{-}(1-\hat{n}_D)$ \\
\hline
$\epsilon_{-}+U_{LR} $ & $-\hat{d}_- \hat{n}_{+}(1-\hat{n}_D)$ & $\hat{d}_- \hat{n}_{+}(1-\hat{n}_D)$  \\
\hline
$\epsilon_{+}+U+U_{LR}$ & $\hat{d}_+ \hat{n}_{-}\hat{n}_D$ & $\hat{d}_+ \hat{n}_{-}\hat{n}_D$ \\
\hline
$\epsilon_{-}+U+U_{LR}$ & $-\hat{d}_- \hat{n}_{+}\hat{n}_D$ & $\hat{d}_- \hat{n}_{+}\hat{n}_D$ \\
\hline
\end{tabular}
\caption{Full-secular jump operators induced by the left ($L$) and right ($R$) reservoirs for each Bohr frequency.}
\label{table:TablaIII}
\end{table}

\section{\label{app:redfieldglobal}Dissipators of the global equation}

The secular approximation assumes that different Bohr frequencies satisfy $|\omega - \omega'| \gg 1/\tau_{S}$, where $\tau_S$ denotes the characteristic relaxation time of the system. Thus, all terms with $\omega\neq\omega'$ in Eq.\eqref{eq:REdfield} oscillate quickly and are averaged out. As a consequence, only the terms with $\omega=\omega'$ are preserved. In the Schrodinger picture, the dissipator associated with the reservoir $r$ is
\begin{align*}
    \mathcal{L}_{r} \hat \rho_{S}(t)= & \sum_{l,\omega} \gamma^{(r)}_{l}(\omega) \hat{A}_{r,l}(\omega) \hat \rho_{S}(t)\hat{A}^{\dagger}_{r,l} (\omega)  \\
     & -\sum_{l,\omega} \frac{\gamma^{(r)}_{l}(\omega)}{2} \{ \hat{A}^{\dagger}_{r,l} (\omega)\hat{A}_{r,l}(\omega),\hat \rho_{S}(t)\}. \nonumber
\end{align*}

In our setup, the relaxation rate is characterized by the tunneling rate $\kappa_L$, such that $1/\tau_S \sim \kappa_L$. We choose the interaction parameters $U$ and $U_{LR}$ such that the corresponding frequency differences, $|\omega - \omega'| \sim U,U_{LR}, U+U_{LR}$ are always much larger than $\kappa_L$. 

Of particular importance is the coherent tunneling strength $g$, which induces Bohr-frequency splittings of order $|\omega - \omega'| \sim g$. 
Two distinct regimes must therefore be distinguished depending on the ratio $g/\kappa_{L}$. When $g/\kappa_L \gg 1$, a full secular approximation can be performed.

For the fermionic reservoirs, the jump operators and its transition frequencies are listed in Table \ref{table:TablaIII}, and the decay rates in Eq.~\eqref{eq:fermirates}. Thus, the corresponding dissipators read

\begin{widetext}
\begin{align*}
   & \mathcal{L}_{B_j}^{\rm (f.s)}  =  \frac{1}{2}\sum_{\sigma \neq \sigma'} ( \gamma_+^{(B_j)}(\epsilon_\sigma)\mathcal{D}[\hat{d}^{\dagger}_{\sigma}(\mathbf{1}-\hat{n}_{D})(\mathbf{1}-\hat{n}_{\sigma' }) ]  + \gamma_-^{(B_j)}(\epsilon_\sigma)\mathcal{D}[\hat{d}_{\sigma}(\mathbf{1}-\hat{n}_{D})(\mathbf{1}-\hat{n}_{\sigma'}) ]   \nonumber \\
                    & + \gamma_+^{(B_j)}(\epsilon_\sigma +U)\mathcal{D}[\hat{d}^{\dagger}_{\sigma}\hat{n}_{D}(\mathbf{1}-\hat{n}_{\sigma'}) ]  +\gamma_-^{(B_j)}(\epsilon_\sigma+U)\mathcal{D}[\hat{d}_{\sigma}\hat{n}_{D}(\mathbf{1}-\hat{n}_{\sigma'}) ]  
                    \\
                   & + \gamma_{+}^{(B_j)}(\epsilon_\sigma+U_{LR})\mathcal{D}[\hat{d}^{\dagger}_{\sigma}(\mathbf{1}-\hat{n}_{D})\hat{n}_{\sigma'} ] + \gamma_{-}^{(B_j)}(\epsilon_\sigma+U_{LR})\mathcal{D}[\hat{d}_{\sigma}(\mathbf{1}-\hat{n}_{D})\hat{n}_{\sigma'} ]  \nonumber \\
                  & + \gamma_{+}^{(B_j)}(\epsilon_\sigma +U+U_{LR})\mathcal{D}[\hat{d}^{\dagger}_{\sigma}\hat{n}_{D}\hat{n}_{\sigma'} ] + \gamma_{-}^{(B_j)}(\epsilon_\sigma +U+U_{LR})\mathcal{D}[\hat{d}_{\sigma}\hat{n}_{D}\hat{n}_{\sigma'} ]) \nonumber , 
\end{align*}
with $j\in \{L,R\}$ and $\sigma,\sigma' = +,-$.
\end{widetext}

Similarly, with the transition frequencies and jump operators summarized in Table \ref{table:TablaII}, the Lindblad dissipator associated with the reservoir $B_D$ can be written as


\begin{widetext}
\begin{equation}
\begin{aligned}
    & \mathcal{L}_{B_D}  = \gamma^{(B_D)}_{+}(\epsilon_{D})\mathcal{D}[\hat{d}^{\dagger}_{D}(\mathbf{1}-\hat{n}_{+})(\mathbf{1}-\hat{n}_{-}) ] + \gamma^{(B_D)}_{-}(\epsilon_{D})\mathcal{D}[\hat{d}_{D}(\mathbf{1}-\hat{n}_{+})(\mathbf{1}-\hat{n}_{-}) ]   \\
                    & + \gamma^{(B_D)}_{+}(\epsilon_{D}+U)\mathcal{D}[\hat{d}^{\dagger}_{D}(\hat{n}_{+}(\mathbf{1}-\hat{n}_{-}) + \hat{n}_{-}(\mathbf{1}-\hat{n}_{+})) ]  \\ 
                    & + \gamma^{(B_D)}_{-}(\epsilon_{D}+U)\mathcal{D}[\hat{d}_{D}(\hat{n}_{+}(\mathbf{1}-\hat{n}_{-}) + \hat{n}_{-}(\mathbf{1}-\hat{n}_{+}))] \\
                   & + \gamma^{(B_D)}_{+}(\epsilon_{D}+2U)\mathcal{D}[\hat{d}^{\dagger}_{D}\hat{n}_{+}\hat{n}_{-} ] + \gamma^{(B_D)}_{-}(\epsilon_{D}+2U)\mathcal{D}[\hat{d}_{D}\hat{n}_{+}\hat{n}_{-} ]. 
\end{aligned}
 \label{app:lindblademon}
\end{equation}
\end{widetext}

The resulting master equation $d_t\hat{\rho}_s=-i[\hat H_S,\hat\rho_S]+{\mathcal L}^{\rm(f.s)}(\hat\rho_S)$ with ${\mathcal L}^{\rm(f.s)}={\mathcal L}_{\rm ph}+{\mathcal L}_{B_D}+{\mathcal L}^{\rm(f.s)}_{B_L}+{\mathcal L}^{\rm(f.s)}_{B_R}$ takes the GKLS form and is called the global Lindblad equation, because the jump operators act globally on the system's eigenstates.

\section{Dissipators of the semilocal equation}
\label{app:appE}
The full secular approximation breaks down when the system exhibits pairs or manifolds of relevant Bohr frequencies that are nearly degenerate, such that $|\omega-\omega'|$ is comparable to or smaller than the relaxation rates. In this situation, the corresponding oscillatory terms do not average out on the dissipative timescale and may contribute significantly to the dynamics.

The partial secular approximation accounts for this by grouping nearly degenerate Bohr frequencies. Frequencies around a value $\omega_q$ that differ by less than the relaxation rate are grouped in the set $x_q$. For frequencies $\omega$ and $\omega'$ within the same $x_q$, one approximates $e^{i(\omega-\omega')t} \simeq 1$.
 Importantly, the variation of $\Gamma_i(\omega)$ for all $\omega$ grouped at the value $\omega_q$, must be negligible, so that $\Gamma_i(\omega)\approx \Gamma_i(\omega_q)$ for all $\omega$ in the group. For frequencies $\omega$ and $\omega'$ in different sets $e^{i(\omega-\omega')t} \simeq 0$, as in the full secular approximation. This procedure leads again to a GKLS master equation, but with jump operators
\begin{equation*}
\hat A_{r,l}(\omega_q)=\sum_{\omega \in x_q} \hat A_{r,l}(\omega),
\end{equation*}
and $\gamma_l^{(r)}(\omega_q)=Re[\Gamma_l^{(r)}(\omega_q)]$ associated with each group. Non-degenerate frequencies correspond to groups containing a single element.

When these collective jump operators can be expressed in terms of operators acting on subsystems of $S$, the resulting generator is commonly referred to as a \emph{semilocal Lindblad equation}. Importantly, this approximation allows coherences within nearly degenerate subspaces to survive and influence the dynamics, while still guaranteeing complete positivity of the evolution.

In the regime $g/\kappa_L \lesssim1$, frequencies differing by 
$g$ cannot be separated. The corresponding Bohr frequencies differing by order of $g$ are therefore grouped into clusters.
\begin{align*}
     (\epsilon_{+},\epsilon_-) & \to \epsilon,  \\
    (\epsilon_{+}+U,\epsilon_-+U) &\to \epsilon + U, \\
    (\epsilon_{+}+U_{LR},\epsilon_- +U_{LR}) & \to \epsilon + U_{LR}, \\
     (\epsilon_{+}+ U+U_{LR},\epsilon_- + U +U_{LR}) & \to \epsilon + U + U_{LR}.
\end{align*}

To obtain the jump operators, we perform the frequency grouping in Eqs.~\eqref{app:jump_op_global1} and \eqref{app:jump_op_global2}, and get

\begin{equation*}
\begin{aligned}
    & e^{i\hat{H}_S t} \hat{d}_{i}e^{-i\hat{H}_S t}  \approx e^{-i\epsilon t} \hat{d}_i (1-\hat{n}_{D})(1-\hat{n}_{j})\\
    & + e^{-i(\epsilon + U_{LR})t}\hat{d}_i (1-\hat{n}_{D})\hat{n}_{j} \\
    & + e^{-i(\epsilon + U)t}\hat{d}_i (1-\hat{n}_{j})\hat{n}_{D} + e^{-i(\epsilon + U + U_{LR})t}\hat{d}_{i}\hat{n}_{D}\hat{n}_{j},
\end{aligned}
\end{equation*}
with $i\neq j$ and $i,j =L,R$. Consequently, the dissipators acting on the $L$ and $R$ dots read
\begin{widetext}
\begin{equation}
\begin{aligned}
   & \mathcal{L}_{B_j}^{\rm(p.s)}  =   \gamma_+^{(B_j)}(\epsilon)\mathcal{D}[\hat{d}^{\dagger}_{j}(\mathbf{1}-\hat{n}_{D})(\mathbf{1}-\hat{n}_{k}) ]  + \gamma_-^{(B_j)}(\epsilon)\mathcal{D}[\hat{d}_{j}(\mathbf{1}-\hat{n}_{D})(\mathbf{1}-\hat{n}_{k}) ]    \\
                    & + \gamma_+^{(B_j)}(\epsilon+U)\mathcal{D}[\hat{d}^{\dagger}_{j}\hat{n}_{D}(\mathbf{1}-\hat{n}_{k}) ]  +\gamma_-^{(B_j)}(\epsilon+U)\mathcal{D}[\hat{d}_{j}\hat{n}_{D}(\mathbf{1}-\hat{n}_{k}) ]  
                    \\ 
                   & + \gamma_{+}^{(B_j)}(\epsilon+U_{LR})\mathcal{D}[\hat{d}^{\dagger}_{j}(\mathbf{1}-\hat{n}_{D})\hat{n}_{k} ] + \gamma_{-}^{(B_j)}(\epsilon+U_{LR})\mathcal{D}[\hat{d}_{j}(\mathbf{1}-\hat{n}_{D})\hat{n}_{k} ]    \\
                  & + \gamma_{+}^{(B_j)}(\epsilon+U+U_{LR})\mathcal{D}[\hat{d}^{\dagger}_{j}\hat{n}_{D}\hat{n}_{k} ] + \gamma_{-}^{(B_j)}(\epsilon+U+U_{LR})\mathcal{D}[\hat{d}_{j}\hat{n}_{D}\hat{n}_{k} ]  , 
\end{aligned}
\label{app:lindbladsemilocallr}
\end{equation}
with $j,k=L,R$, and $j\neq k$.\\
\end{widetext}

\begin{figure}[h!]
\includegraphics[scale = 1]{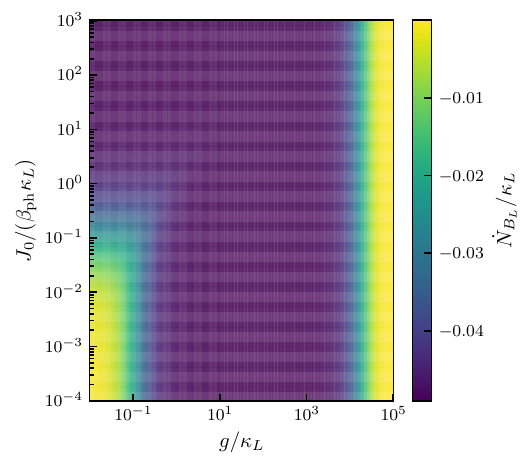}
\caption{\label{fig:appN3D}  Plot of the particle current $\dot{N}_{B_L}/\kappa_L$ as a function of $g/\kappa_L$ and $J_0/(\beta_{\mathrm{ph}} \kappa_L)$. From $g/\kappa_L \lesssim 10^{1}$, data was obtained with the semilocal equation. From $g/\kappa_L \gtrsim 10^{1}$, data was obtained with the secular equation. Parameters: All parameters are the same of Fig.~\ref{fig:lindbladcomp1}, except for $g/ \kappa_{L}$ and $J_0/(\beta_{\mathrm{ph}} \kappa_L)$.}
\end{figure}

The Lindblad dissipator of the demon $D$ is given by Eq.~\eqref{app:lindblademon}. Using the identities  $\hat{n}_+ + \hat{n}_{-} = \hat{n}_L + \hat{n}_R$ and $\hat{n}_+ \hat{n}_- = \hat{n}_L \hat{n}_R$, the corresponding jump operators of this dissipator can be written in terms of local operators.

From the explicit semilocal form of the jump operators in Eqs.~\eqref{app:lindbladsemilocallr}, and \eqref{app:lindblademon}, the resulting master equation, $\mathcal{L}^{(\mathrm{p.s})} = \mathcal{L}_{B_L}^{(\mathrm{p.s})} + \mathcal{L}_{B_R}^{(\mathrm{p.s})}+ \mathcal{L}_{B_D} + \mathcal{L}_{\mathrm{ph}}$, is referred to as a semilocal Lindblad equation. 

Transitions that originally correspond to distinct energy exchanges at the Bohr frequencies $\epsilon_+$ and $\epsilon_-$ are now described by a single effective process associated with the energy $\epsilon$. This loss of energy resolution, at the scale $g$ must be accompanied with a consistent thermodynamic bookkeeping reflecting the frequency grouping. In particular, this motivates the introduction of an effective thermodynamic Hamiltonian 
\begin{equation*}
    \hat{H}_{TD} = \hat{H}_{S} - g(\hat{d}^{\dagger}_L \hat{d}_{R} + \hat{d}^{\dagger}_{R}\hat{d}_{L} ),
\end{equation*}
which satisfies $[\hat{H}_{TD},\hat{A}_{r,l}(\omega_q)]=\omega_q\hat{A}_{r,l}(\omega_q)$, with $\omega_q$ the grouped Bohr frequencies and $\hat{A}_{r,l}(\omega_q)$ the jump operators of the partial secular approximation listed in Tables \ref{table:TablaII} and \ref{table:TablaIV}.

\begin{table}[h]
\centering
\begin{tabular}{|c|c|c|}
\hline
\textbf{Frequency} & \textbf{L} & \textbf{R} \\
\hline
$\epsilon$ & $\hat{d}_L (1-\hat{n}_{D})(1-\hat{n}_{R})$ & $\hat{d}_R (1-\hat{n}_{D})(1-\hat{n}_{L})$ \\
\hline
$\epsilon+U$ & $\hat{d}_L (1-\hat{n}_{R})\hat{n}_D$ &  $\hat{d}_R (1-\hat{n}_{L})\hat{n}_D$ \\
\hline
$\epsilon+U_{LR} $ & $\hat{d}_L \hat{n}_{R}(1-\hat{n}_D)$ & $\hat{d}_R \hat{n}_{L}(1-\hat{n}_D)$ \\
\hline
$\epsilon+U+U_{LR}$ & $\hat{d}_L \hat{n}_{R}\hat{n}_D$ & $\hat{d}_R \hat{n}_{L}\hat{n}_D$ \\
\hline
\end{tabular}
\caption{Partial-secular jump operators induced by the left ($L$) and right ($R$) reservoirs for each Bohr frequency.}
\label{table:TablaIV}
\end{table}

 This thermodynamic Hamiltonian defines the reference energy structure used to compute heat and energy currents. Under this approximation, the thermodynamic description is equivalent to those of~\cite{potts2021thermodynamically,trushechkin2021unified} and coincides with that of~\cite{hewgill2021quantum}.

\section{\label{app:energy3d}Maxwell-demon criterion and engine regime}



In the main text, we present Fig.~\ref{fig:ratio}
showing that \(|\dot E_1/T\dot I_1|\lesssim 10^{-1}\) for all values that $g/\kappa_L$ and $J_0/(\beta_{\rm ph}\kappa_L)$ take along the quantum-to-classical transitions. 

Here we present a complementary plot, Figure~\ref{fig:appN3D}, confirming that the \(LR\) subsystem extracts work by transporting particles against the bias, \(\dot N_{B_L}<0\), for all values that $g/\kappa_L$ and $J_0/(\beta_{\rm ph}\kappa_L)$ take across the parameter region spanning the two quantum-to-classical crossovers studied in this work.  


\bibliography{apssamp}

\end{document}